\documentclass[journal]{IEEEtran}
\usepackage[cmex10]{amsmath,empheq}
\usepackage{algorithmic}
\usepackage{algorithm}
\usepackage{array}
\usepackage[caption=false,font=normalsize,labelfont=sf,textfont=sf]{subfig}
\usepackage{textcomp}
\usepackage{stfloats}
\usepackage{url}
\usepackage{verbatim}
\usepackage{graphicx}
\usepackage{amssymb}
\usepackage{cite}
\usepackage{bm}
\usepackage{xcolor}
\usepackage[nolist]{acronym}
\usepackage{comment}
\usepackage{hyperref}
\usepackage{booktabs}
\usepackage{tikz}
\usetikzlibrary{arrows.meta, positioning, shapes.multipart, calc}

\begin{document}
\newcommand{\hop}{\mathsf{H}}
\newcommand{\bo}{\boldsymbol}
\newcommand{\complex}{\mathbb C}
\newcommand{\comp}{\mathbin{\circ}}
\newcommand{\Fro}{\mathrm{F}}

\begin{acronym}

\acro{5G}{fifth generation}
\acro{5G-A}{fifth generation-advanced}
\acro{6G}{sixth generation}
\acro{3GPP}{3rd generation partnership project}
\acro{TS}{technical specification}
\acro{MIMO}{multiple-input multiple-output}
\acro{MU}{multi-user}
\acro{MU}{multi-user}
\acro{BS}{base station}
\acro{gNB}{Next-Generation Node B}
\acro{UE}{user equipment}
\acro{UL}{uplink}
\acro{DL}{downlink}
\acro{RE}{resource element}
\acro{DM-RS}{demodulation reference symbol}
\acro{CDM}{code division multiplex}
\acro{PSK}{phase-shift keying}
\acro{QAM}{quadrature amplitude modulation }

\acro{AI}{artificial intelligence}
\acro{ML}{machine learning}
\acro{NN}{neural network}
\acro{DNN}{deep neural network}
\acro{FNN}{feedforward neural network}
\acro{FCNN}{fully connected neural network}
\acro{CNN}{convolutional neural network}
\acro{RL}{reinforcement learning}
\acro{GNN}{graph neural network}

\acro{SNR}{Signal-to-Noise Ratio}
\acro{UMi}{urban Micro}
\acro{UMa}{urban Macro}
\acro{RF}{radio frequency}
\acro{AWGN}{additive white Gaussian noise}
\acro{OFDM}{orthogonal frequency-division multiplexing}
\acro{CSI}{channel state information}

\acro{DFT}{discrete Fourier transform}
\acro{IDFT}{inverse discrete Fourier transform}
\acro{CP}{cyclic prefix}
\acro{PA}{power amplifier}
\acro{PAPR}{peak-to-average power ratio}
\acro{ACLR}{adjacent channel leakage ratio}

\acro{SI}{superimposed}
\acro{SIP}{superimposed pilot}
\acro{SID}{superimposed data}
\acro{DP}{data-to-pilot}
\acro{DMRS}{demodulation reference signal}

\acro{OCC}{orthogonal cover code}
\acro{ROT}{rotational}

\acro{DeepTx}{deep learning transmitter}
\acro{DeepRx}{deep learning receiver}
\acro{MixedDeepTRX}{Mixed deep learning transceiver}

\acro{LS}{least squares}
\acro{LMMSE}{linear minimum mean square error}
\acro{MRC}{maximum ratio combining}

\acro{ULA}{uniform linear array}

\acro{RAN}{radio access network}
\acro{SVD}{singular value decomposition}
\acro{JCDD}{joint channel estimation, detection, and
decoding}
\acro{IC}{interference cancellation}
\acro{TD}{time-domain}
\acro{FD}{frequency-domain}
\acro{BER}{bit error rate}
\acro{BLER}{block error rate}

\end{acronym}

\title{Superimposed DMRS for Spectrally Efficient 6G Uplink Multi-User OFDM: Classical vs AI/ML Receivers}
\author{Sajad Rezaie, Mikko Honkala, Dani Korpi, Dick Carrillo Melgarejo, Tomasz Izydorczyk, Dimitri Gold, and Oana-Elena Barbu
\thanks{Sajad Rezaie is with Nokia, Aalborg, Denmark. (e-mail: sajad.rezaie@nokia.com)}
\thanks{Mikko Honkala is with Nokia Bell Labs, Espoo, Finland. (e-mail: mikko.honkala@nokia-bell-labs.com)}
\thanks{Dani Korpi is with Nokia Bell Labs, Espoo, Finland. (e-mail: dani.korpi@nokia-bell-labs.com)}
\thanks{Dick Carrillo Melgarejo is with Nokia, Espoo, Finland (e-mail: dick.carrillo\_melgarejo@nokia.com) and LUT university (dick.carrillo.melgarejo@lut.fi)}
\thanks{Tomasz Izydorczyk is with Nokia, Wroclaw, Poland. (e-mail: tomasz.izydorczyk@nokia.com)}
\thanks{Dimitri Gold is with Nokia, Espoo, Finland. (e-mail: dimitri.gold@nokia.com)}
\thanks{Oana-Elena Barbu is with Nokia, Aalborg, Denmark. (e-mail: oana-elena.barbu@nokia.com)}

\thanks{Manuscript received ???; revised ???.}}


\maketitle

\begin{abstract}
Fifth-generation (5G) systems utilize orthogonal demodulation reference signals (DMRS) to enable channel estimation at the receiver. These orthogonal DMRS—also referred to as pilots—are effective in avoiding pilot contamination and interference from both the user’s own data and that of others. However, this approach incurs a significant overhead, as a substantial portion of the time-frequency resources must be reserved for pilot transmission. Moreover, the overhead increases with the number of users and transmission layers.

To address these limitations in the context of emerging sixth-generation (6G) systems and to support data transmission across the entire time-frequency grid, the superposition of data and DMRS symbols has been explored as an alternative DMRS transmission strategy. In this study, we propose an enhanced version of DeepRx, a deep convolutional neural network (CNN)-based receiver, capable of estimating the channel from received superimposed (SI) DMRS symbols and reliably detecting the transmitted data. We also design a conventional receiver for comparison, which estimates the channel from SI DMRS using classical signal processing techniques. Extensive evaluations in both uplink single-user and multi-user scenarios demonstrate that DeepRx consistently outperforms the conventional receivers in terms of performance.
\end{abstract}


\section{Introduction}
\IEEEPARstart{O}{ver} the past few years, the wireless communications industry has been progressively and steadily adopting \ac{ML} techniques for the design of the physical layer in \ac{5G-A} and \ac{6G} \acp{RAN} \cite{hoydis17}. Given that these procedures require coordination between the radio nodes of different vendors, \ac{3GPP} is studying and standardizing the application of \ac{ML} for multiple use-cases in Release 19. These include beam management, positioning, \ac{CSI} feedback, and mobility \cite{TR38843, TR38744}. Looking ahead, future \ac{6G} networks are envisioned to increasingly rely on a fully \ac{AI}-native air interface, where \ac{ML} models are tightly integrated into the physical layer to enhance flexibility, generalization, and performance.

Among the various \ac{ML}-based solutions for the physical layer, neural receivers have emerged as a promising approach \cite{9593152, korpi21, 10051898, qi23}.  The concept has since been popularized under the name \ac{DeepRx} \cite{Honkala21} and has undergone different implementations. The original \ac{DeepRx} and its variant called HybridDeepRx \cite{Jaakko21}  departs from the conventional sequential processing pipeline of the \ac{OFDM} receiver involving channel estimation, equalization, detection and decoding. Instead, they adopt holistic neural receiver architectures based on \acfp{CNN} which directly infer transmitted symbols without explicitly estimating the channel or performing traditional equalization.

Further enhancements to DeepRx have been proposed to support transmissions using waveforms alternative to OFDM. For instance, \cite{qi23} introduces a non-trainable \ac{IDFT}-deprecoder block intercalated with the existing convolutional layers of DeepRx. This modification enables the network to better exploit the frequency spreading characteristics of single-carrier transmission more effectively.

More recently,\cite{Cammerer23} presents a neural receiver tailored for \ac{MU}  \ac{MIMO} systems, which is claimed to be fully \ac{5G} compliant. This design employs \acp{GNN} to handle a variable number of users without retraining, while jointly performing channel estimation, equalization, and demapping for each user's \ac{UL} signal.  

Beyond neural receivers alone, other studies have extended \ac{ML} to the joint design of wireless transmitters and receivers \cite{faycal22, faycal22TC, goutay21, korpi23, 9681990, 9394761}. For example, \cite{faycal22} proposes two \ac{ML}-based transceiver variants: the first variant enables a neural transmitter to learn and share an irregular complex constellation, with a matched neural receiver tailored to demap the symbols drawn from this constellation. The second variant employs \acf{SI} pilot transmission where pilot symbols are transmitted on top of the data symbols on the same resources \cite{8497042, 10507003, 9850397}. In this study, a neural receiver is show to effectively leverage the pilot-data superposition for data recovery.

In \cite{AI_Li25}, \ac{IC}-based \ac{JCDD} receivers are proposed for processing \ac{SI} pilots transmissions in single-user MIMO-OFDM systems. Their design applies \ac{ML}-driven channel estimators  and demonstrates improved robustness in dynamic channel conditions compared to traditional \ac{LMMSE} estimators. Similarly, \cite{Interference_Han25} proposes an \ac{IC}-based neural receiver for \ac{SI} pilots in downlink multi-layer transmissions. This design utilizes  \ac{DL}-based precoding methods such as \ac{SVD}, which enables nearly parallelized channels across layers, thus reducing inter-layer interference in the received \ac{SI} pilots. In a more pioneering approach, the authors of \cite{korpi23} extended \ac{ML}-based transceiver designs to support pilotless spatial multiplexing transmissions in \ac{MIMO} systems, demonstrating 15-20\% spectral efficiency gains. Likewise, \cite{9681990} presents a learnable transceiver that jointly optimizes the pulse and constellation shapes.

\begin{figure*}[!t]
\centering
\scalebox{0.65}{
\begin{tikzpicture}[
  block/.style={draw, minimum width=2cm, minimum height=1.2cm, align=center},
  arrow/.style={-Latex, thick},
  every node/.style={font=\small}
]

\node[block] (bits1)                          {Encode};
\node[block, right=1cm of bits1] (mod1)     {Modulate};
\node[block, right=1cm of mod1] (dmrs1)     {DMRS Insertion \\(if any)};
\node[block, right=1cm of dmrs1] (resource1) {Layer and \\ Resource Mapping};
\node[block, right=1cm of resource1] (ofdm_mod1) {OFDM \\ Mod};
\node[above=2pt of bits1] {\large $\mathbf{u}^{(1)}$};

\draw[arrow] (bits1) -- node[above] {\large $\mathbf{b}^{(1)}$} (mod1);
\draw[arrow] (mod1) -- node[above] {\large $\mathbf{d}^{(1)}$} (dmrs1);
\draw[arrow] (dmrs1) -- node[above] {\large $\mathbf{x}^{(1)}$} (resource1);
\draw[arrow] (resource1) -- node[above] {} (ofdm_mod1);

\node[block, below=1.8cm of bits1] (bits2)    {Encode};
\node[block, right=1cm of bits2] (mod2)     {Modulate};
\node[block, right=1cm of mod2] (dmrs2)     {DMRS Insertion \\(if any)};
\node[block, right=1cm of dmrs2] (resource2) {Layer and \\ Resource Mapping};
\node[block, right=1cm of resource2] (ofdm_mod2) {OFDM \\ Mod};
\node[above=2pt of bits2] {\large $\mathbf{u}^{(K)}$};

\draw[arrow] (bits2) -- node[above] {\large $\mathbf{b}^{(K)}$} (mod2);
\draw[arrow] (mod2) -- node[above] {\large $\mathbf{d}^{(K)}$} (dmrs2);
\draw[arrow] (dmrs2) -- node[above] {\large $\mathbf{x}^{(K)}$} (resource2);
\draw[arrow] (resource2) -- node[above] {} (ofdm_mod2);

\node at ($(bits2)!0.5!(mod2)+(0,1.5)$) {\Large $\vdots$};

\node[block, right=1cm of ofdm_mod1] (chan1)     {MIMO Channel \\ $\mathbf{H}^{(1)}$};
\draw[arrow] (ofdm_mod1) -- node[above] {} (chan1);

\node[block, right=1cm of ofdm_mod2] (chan2)     {MIMO Channel \\$\mathbf{H}^{(K)}$};
\draw[arrow] (ofdm_mod2) -- node[above] {} (chan2);

\node[circle, draw, minimum size=0.9cm, right=2.8cm of chan1] (sum) {+};
\node[block, below=1.4cm of sum] (noise) {AWGN};
\node[above = 0.4cm of noise, xshift=0.3cm] {\large $\mathbf{n}$};
\draw[arrow] (noise.north) -- (sum.south);

\draw[arrow] (chan1) -- (sum);
\draw[arrow] (chan2) -- (sum);

\node[block, right=1cm of sum] (recv) {OFDM \\ DeMod};
\node[above=2pt of recv] {\large $\mathbf{y}$};
\draw[arrow] (sum) -- (recv);

\end{tikzpicture}}
\caption{Block diagram of the considered UL MU-MIMO OFDM system.}
\label{fig:mu_mimo_ofdm}
\end{figure*}
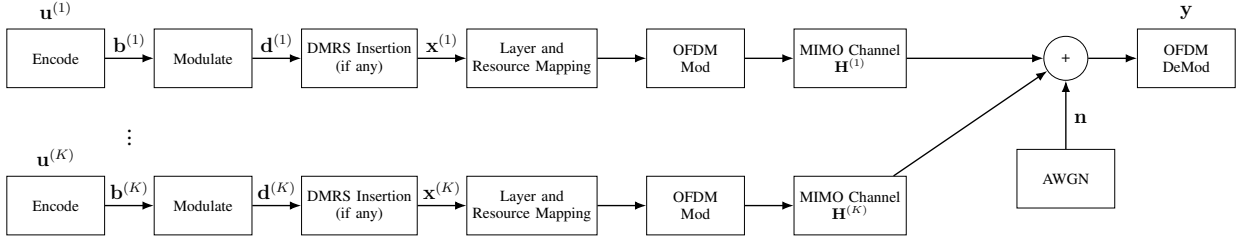

To further leverage the flexibility of ML tools, several works have addressed the design of \ac{ML}-based transceivers for processing \ac{OFDM} signals affected by various forms of self-interference or hardware-induced nonlinearities \cite{faycal21, 10017176, Jaakko21, 9834021, goutay21}. For example, \cite{faycal21} proposes a neural receiver capable of operating in \ac{CP}-less and pilotless \ac{OFDM} systems. Another notable work is \cite{Jaakko21}, which introduces a neural receiver designed to handle signals impaired by the \ac{PA} nonlinearities. This work is extended in \cite{9834021}, where phase noise is additionally considered during deep receiver training. Furthermore, \cite{goutay21} proposes an end-to-end waveform learning framework in which a deep transmitter learns an irregular constellation under \ac{PAPR} and \ac{ACLR} constraints.

In this contribution, we discuss how \ac{SI} \ac{DMRS} can be configured to support multi-user MIMO-OFDM transmissions, and present the criteria for \ac{SI} \ac{DMRS} patterns to allow proper channel estimation at the receiver side. 
Following the paradigm of waveform learning, in this paper we propose DeepRx, a deep \ac{CNN}-based receiver capable of processing received signals in \ac{UL} \ac{MU}-\ac{MIMO} systems with \ac{SI} \ac{DMRS} scheme. 
In addition, we design an iterative receivers that can handle different types of \ac{SI} \ac{DMRS} configurations in \ac{UL} multi-user scenarios. 
We provide a comprehensive collection of numerical results for various MIMO configurations, demonstrating that the proposed DeepRx network outperforms the classical receivers in all the scenarios with \ac{SI} \ac{DMRS} transmission. 
Moreover, the proposed DeepRx offers throughput gain with \ac{SI} \ac{DMRS} transmission compared to the DeepRx with \ac{5G} orthogonal \ac{DMRS}.

The rest of this paper is organized as follows. In Section \ref{section:System_Model}, we describe the considered system model and the possible \ac{DMRS} transmission schemes. Section \ref{section:Design_SI_DMRS} provides detailed information on how \ac{SI} \ac{DMRS} can be designed to ensure proper channel estimation at the receiver. We propose classical and ML-based receivers for decoding data with \ac{SI} \ac{DMRS} transmission in Section \ref{section:Receiver_SI_DMRS}. Section \ref{section:Evaluation} presents numerical results for the proposed classical and ML-based receivers. The conclusions are presented in Section \ref{section:Conclusions}.

\textit{Notation} In this paper,  $a$, $\boldsymbol{a}$, and $\boldsymbol{A}$ denote a scalar, a vector, and a matrix, respectively. Transposition and complex transposition of vectors and matrices are respectively represented as $(\cdot)^T$ and $(\cdot)^H$. In addition, $\boldsymbol{A}^{(k)}$ denotes a matrix associated with the k-th transmission, while $\boldsymbol{A}^{(k, u)}$ denotes a matrix of the k-th transmission at u-th iteration. $\mathcal{A}$ denotes a finite set, with $A$ being its cardinality.

\section{System Model and Transmission Schemes}\label{section:System_Model}
In this paper, we consider a \ac{MU}-\ac{MIMO} \ac{UL} scenario comprising $K$ \acp{UE} and a single \ac{BS}. Without loss of generalization, we assume that both the \ac{BS} and \acp{UE} are equipped with \acp{ULA}, where the $k$-th \ac{UE} and the \ac{BS} have $N^{(k)}_T$ and $N_R$ antenna elements, respectively. 

We consider an \ac{OFDM} system with $n_F$ subcarriers and $n_T$ OFDM symbols as shown in \ref{fig:mu_mimo_ofdm}. After removing the \ac{CP} and applying \ac{DFT}, the received signal at the \ac{BS} on the $i$-th subcarrier and $j$-th \ac{OFDM} symbol can be represented by:
\begin{equation}
   \mathbf{y}_{i, j} = \sum_{k=1}^{K} \mathbf{H}^{(k)}_{i, j} \mathbf{x}^{(k)}_{i, j} + \mathbf{n}_{i,j},
   \label{eq:y_ij}
\end{equation}
where $\mathbf{H}^{(k)}_{i, j} \in \mathcal{C}^{N_R \times N^{(k)}_T}$ represents the \ac{MIMO} channel matrix between the $k$-th \ac{UE} and \ac{BS} at subcarrier $i$ and symbol $j$, $\mathbf{x}^{(k)}_{i,j} \in \mathcal{C}^{N^{(k)}_T \times 1}$ denotes the transmitted signal vector from the $k$-th UE \footnote{ During the transmission on the $j$-th OFDM symbol, a vector $\mathbf{u}_{j}^{(k)}$ of information bits for the $k$-th UE is encoded and interleaved from a coded bit sequence
$\mathbf{b}_{j}^{(k)}$. The sequence is then modulated and mapped to resource elements across the available transmission layers, resulting in the symbol vectors $\mathbf{x}^{(k)}_{i, j}$.}, and $\mathbf{n}_{i,j} \in \mathcal{C}^{N_R \times 1}$ is additive white complex Gaussian noise vector at the receiver. Note that each UE may transmit up to $N^{(k)}_T$ independent data streams, also referred as transmission layers.

\subsection{DMRS Transmission Schemes}
\Acp{DMRS}, which are known reference signals transmitted across all \ac{MIMO} layers, are fundamentally utilized by the \ac{BS} for accurate estimation of the channel responses $\mathbf{H}^{(k)}_{i,j}$. This is a critical prerequisite for reliable data demodulation. The \ac{BS} adaptively configures the time-frequency positions and sequence values of these \acp{DMRS} based on key system parameters, including the number of active users, the number of transmission layers and antenna ports, and the prevailing channel conditions.

In the most general case, the transmitted symbols by the $k$-th \ac{UE} at the $i$th subcarrier and $j$th \ac{OFDM} symbol can be expressed as
\begin{equation}
    \mathbf{x}^{(k)}_{i,j} = \sqrt{1-\mathcal{E}^{(k)}_{i,j}}\mathbf{d}^{(k)}_{i,j} + \sqrt{\mathcal{E}^{(k)}_{i,j}}{\mathbf{p}}^{(k)}_{i,j},
    \label{eq:x_ij}
\end{equation}
where $\mathcal{E}^{(k)}_{i,j}$ denotes the ratio in between the power of pilot/DMRS $\mathbf{p}^{(k)}_{i,j}$ and transmitted symbol $\mathbf{x}^{(k)}_{i,j}$. At the $k$-th \ac{UE}, the data symbols $\mathbf{d}^{(k)}_{i,j}$ are drawn from a modulation constellation $\mathcal{M}^{(k)}$ consisting of $M^{(k)}$ constellation points. Typically, these constellations are regular and follow closed-form definitions such as \ac{QAM} or \ac{PSK}.
\begin{figure*}
    \centering
    \includegraphics[width=0.95\linewidth]{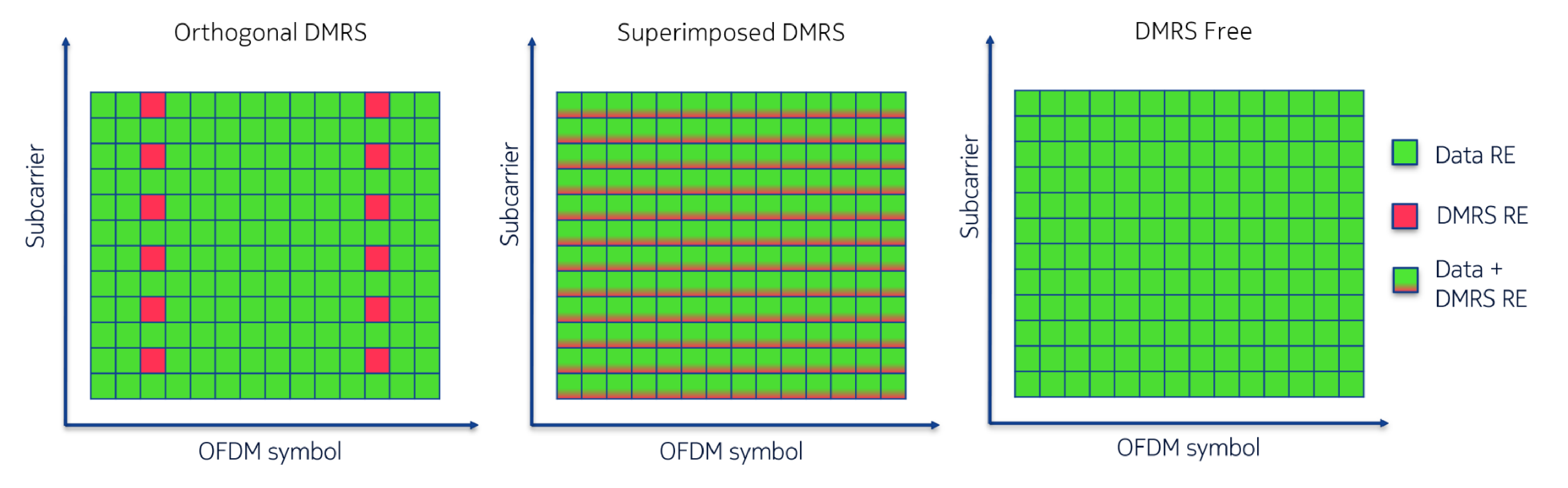}
    \caption{ DMRS transmission schemes considered in this paper.}
    \label{fig:resource_grids}
\end{figure*}

In this study, as illustrated in \autoref{fig:resource_grids}, we consider two \ac{DMRS} transmission schemes alongside a baseline scheme:
\begin{itemize}
    \item \textit{Orthogonal \ac{DMRS} transmission}: In this scheme, disjoint \ac{DMRS} and data \acp{RE} are used, corresponding to $\mathcal{E}^{(k)}_{i,j} = 1$ for DMRS and $\mathcal{E}^{(k)}_{i,j} = 0$ for data. In \ac{5G}, a subset of the time-frequency \acp{RE} is exclusively allocated for \ac{DMRS} transmission, thereby ensuring orthogonality between data and \ac{DMRS} symbols. Moreover, \ac{DMRS} sequences are designed to be orthogonal across both inter- and intra-\ac{UE} transmission layers. The \ac{BS} estimates the channel response for each communication layer based on the received signal at the respective \ac{DMRS} \acp{RE}.
    
    \item \textit{\Acf{SI}-\ac{DMRS} transmission}: In this scheme, some or all transmitted symbols $\mathbf{x}^{(k)}_{i,j}$ comprise a superposition of both data and \ac{DMRS} components, where the \ac{DMRS} power fraction is defined by $\mathcal{E}^{(k)}_{i,j} \ll 1$. Due to this superposition, the scheme is referred to as \ac{SI}-\ac{DMRS} transmission. Pure data symbols, if present, correspond to $\mathcal{E}^{(k)}_{i,j} = 0$.
    
    \item \textit{\ac{DMRS}-Free transmission}: To enhance spectral efficiency, \acp{UE} may allocate their entire transmit power exclusively to data symbols. However, when employing conventional constellations such as \ac{PSK} or \ac{QAM}, a receiver may not be capable of successfully estimating the channel response and recovering transmitted information. 
    Enabling \ac{DMRS}-Free channel estimation at \ac{BS} requires transmission using irregular constellation. Although \Acl{ML} methods have been proposed to learn proper irregular constellations, scaling these solutions to large MIMO systems remains a significant challenge.
\end{itemize}
In this work, we restrict our analysis to regular \ac{QAM} constellations and treat the \ac{DMRS}-Free transmission scheme as an upper bound on achievable throughput.

\section{Superimposed DMRS Design for MU-MIMO OFDM Systems} \label{section:Design_SI_DMRS}
In \ac{5G}, orthogonal \ac{DMRS} transmission enables the receiver to estimate the channel for each transmission layer without suffering from inter-layer pilot contamination or cross \ac{DP} interference originating from the same or other layers. However, allocating dedicated time-frequency resources exclusively  for \ac{DMRS} transmission reduces the effective data throughput.

Alternatively, superimposing \ac{DMRS} symbols on top of data symbols permits data transmission over all resource elements, thereby maximizing spectral efficiency and  link throughput. Nevertheless,the presence of intra- and inter-layer interference from \ac{DMRS} and data components renders accurate channel estimation more challenging in \ac{SI} \ac{DMRS} transmission scheme. 

Since data occupy the entire time-frequency resource grid in the \ac{SI} \ac{DMRS} scheme, intra- and inter-layer data interference during channel estimation is unavoidable. However, inter-layer interference from the superimposed \ac{DMRS} components can be nearly eliminated by enforcing orthogonality among the configured \ac{SI} \ac{DMRS} patterns or sequences. This orthogonality can be achieved in time, frequency, or code domain, or via a combination of thereof.

In this study, we employ \acp{OCC} to guarantee the orthogonality in code domain among the \ac{DMRS} components of the configured \ac{SI} \ac{DMRS} for different \ac{UE}s/layers. By assigning distinct \ac{OCC}s to each \ac{UE}, their respective reference signals remain mutually orthogonal, as the inner product of two \ac{OCC} sequences is zero. As explained before, this orthogonality effectively suppresses interference between reference signals, enabling the receiver to obtain more precise channel state information estimates.

\begin{figure*}
    \centering
    \includegraphics[width=1\linewidth, trim=10pt 0 15pt 0, clip]{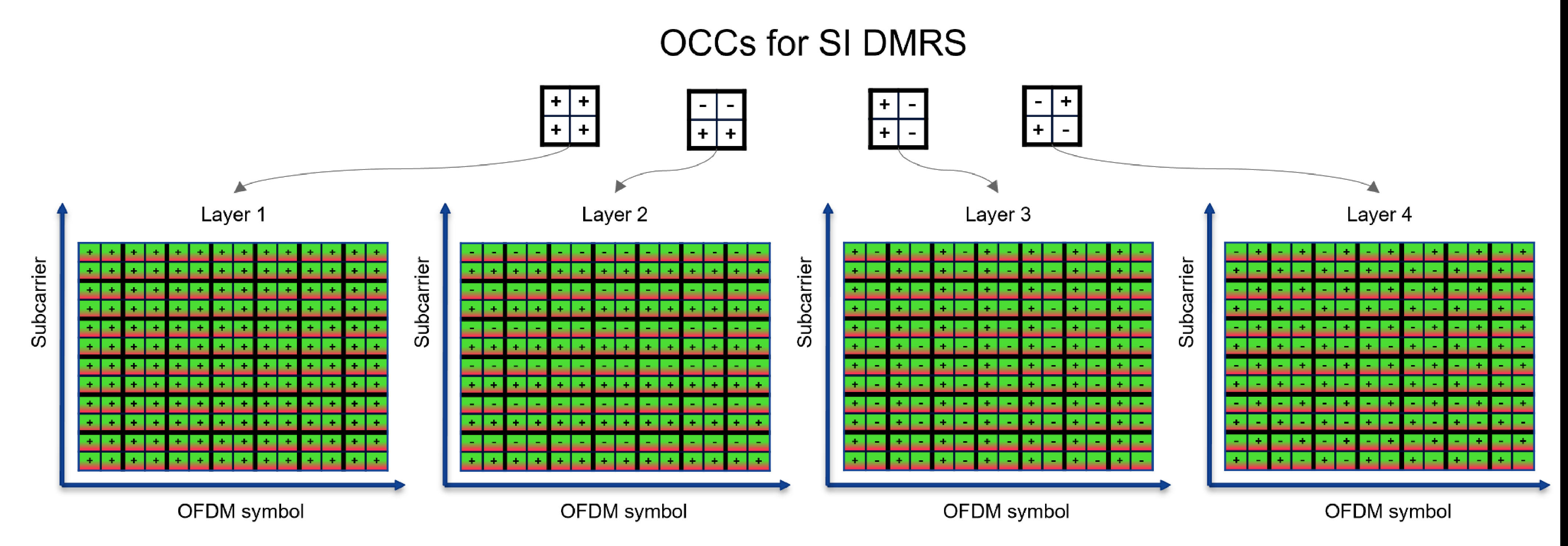}
    \caption{An example of SI DMRS using OCC configuration for four transmission layers. Here, time- and frequency-domain OCCs are configured to ensure orthogonality between SI DMRS of the transmission layers.}
\label{fig:OCC_SI_DMRS_4Layers}
\end{figure*}

It is important to note that \acp{OCC} can be applied in \ac{TD}, \ac{FD}, or both simultaneously. Each OCC sequence assigned to layer/user is spread over the entire allocated time-frequency resource grid, ensuring the preservation of orthogonality among \ac{SI} \ac{DMRS} across the allocated resources. The simplest spreading approach involves repeating  a pre-configured  base \ac{OCC} sequence in time and frequency domains. In addition, to enforce zero-mean of \ac{SI} \ac{DMRS} signals, the repeated \acp{OCC} are multiplied by a zero-mean random sequence, which is identical across all layers. Note that the same random number is used for all the \acp{RE} in an \ac{OCC} group.\autoref{fig:OCC_SI_DMRS_4Layers} illustrates an example of configuring four \acp{OCC} for \ac{SI} \ac{DMRS} in a four-layer transmission scenario.

\section{Receivers for MU-MIMO OFDM}\label{section:Receiver_SI_DMRS}

\subsection{SI DMRS transmission with OCC configuration}
In \ac{SI} \ac{DMRS} transmission scheme, all the \acp{RE} are allocated for both data and pilot transmission, where usually the pilots are transmitted at a lower power level compared to data symbols.
To allow channel estimation at the receiver, pilot symbols transmitted by each user should remain orthogonal to the pilot symbols of all the other users. Therefore, we propose novel classical and AI/ML receivers, exploiting the orthogonality property between \ac{SI} \ac{DMRS} of different transmission layers/users.

\subsubsection{Least-square channel estimation with SI DMRS transmission}
\label{sec:classic_sip_chest}

By plugging Eq. (\ref{eq:x_ij}) in (\ref{eq:y_ij}), the received signal after OFDM demodulation reads
\begin{equation}
   \mathbf{y}_{i, j} = \sum_{k=1}^{K} \mathbf{H}^{(k)}_{i, j}\left(\sqrt{1-\mathcal{E}^{(k)}_{i,j}}\mathbf{d}^{(k)}_{i,j} + \sqrt{\mathcal{E}^{(k)}_{i,j}}{\mathbf{p}}^{(k)}_{i,j}\right) + \mathbf{n}_{i,j}.
   \label{eq:y_ij_sip}
\end{equation}
A \ac{LS} channel estimator that neglects the interference targets to find the channel between the \ac{BS} and the $k$-th user at the $i$-th subcarrier and $j$-th \ac{OFDM} symbol
\begin{equation}
 \hat{\mathbf{H}}_{i,j}^{(k)} = \arg\min_{\mathbf{H}_{i,j}^{(k)}} \left\| \mathbf{y}_{i,j} - \mathbf{H}_{i,j}^{(k)} \sqrt{\mathcal{E}^{(k)}_{i,j}}{\mathbf{p}}^{(k)}_{i,j} \right\|_2^2
   \label{eq:LS_chan_k}
\end{equation} and yields: 

\begin{align}
\mathbf{\hat{H}}_{i,j}^{(k)} &= \frac{ \mathbf{y}_{i,j} \left( \mathbf{p}_{i,j}^{(k)} \right)^H }{ \sqrt{\mathcal{E}^{(k)}_{i,j}} } = \mathbf{H}^{(k)}_{i,j} \nonumber \\
&+  \sum_{\substack{v=1 \\ v \neq k}}^{K} \mathbf{H}^{(v)}_{i,j} \left( \sqrt{\frac{1 - \mathcal{E}^{(v)}_{i,j}}{\mathcal{E}^{(k)}_{i,j}}} \, \mathbf{d}^{(v)}_{i,j}  \right) \left( \mathbf{p}^{(k)}_{i,j} \right)^H \nonumber \\
&\quad + \mathbf{n}_{i,j} \frac{\left( \mathbf{p}^{(k)}_{i,j} \right)^H}{\sqrt{\mathcal{E}^{(k)}_{i,j}}},
\label{eq:LS_chan_k_closed_form}
\end{align}
where the \acf{DP} interference is captured in the second term of the expression. To reduce the \ac{DP} interference plus noise, further smoothening in time-frequency-space domain may be applied to the above LS estimates. Therefore, applying a sliding window with size $(W_f, W_t)$ over the estimated LS results in
\begin{align}
    \mathbf{\hat{{H}}}_{i,j}^{(k)} &= \frac{\sum_{m=-W_f/2+1}^{W_f/2}\sum_{n=-W_t/2+1}^{W_t/2}  \mathbf{\hat{H}}_{i+m,j+n}^{(k)}}{W_f W_t}.
\label{eq:LS_chan_window}
\end{align}
Although considering a larger window sizes reduces the \ac{DP} interference, it introduces a blurring effect. This highlights a trade-off between \ac{DP} interference plus noise reduction and time-frequency resolution.

\begin{figure*}[!t]
\centering
\includegraphics[width=0.7\textwidth,trim=0 0 0 0, clip]{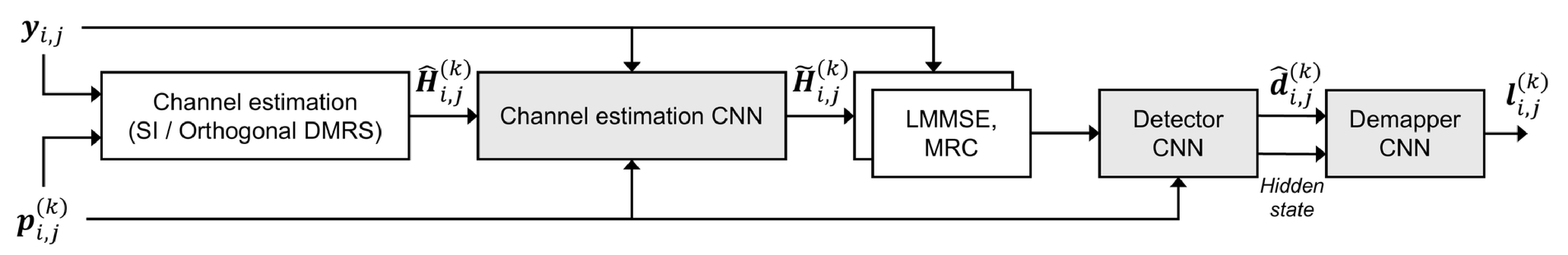}
\caption{A high-level depiction of the DeepRx architecture, used both for superimposed and orthogonal DMRS.}
\label{fig:deeprx_block_diag}
\end{figure*}
\subsubsection{Iterative channel estimation, detection, and decoder with SI DMRS transmission}

To estimate the data symbols from the received signal (\ref{eq:y_ij_sip}), the contribution of the pilot symbols should be first removed: using the channel estimates from (\ref{eq:LS_chan_k_closed_form}), the resulting data signal after pilot removal then reads

\begin{equation}
   \mathbf{\tilde{y}}_{i, j} =  \mathbf{y}_{i, j}- \sum_{k=1}^{K} \mathbf{\hat{H}}^{(k)}_{i, j}\sqrt{\mathcal{E}^{(k)}_{i,j}}{\mathbf{p}}^{(k)}_{i,j}.
   \label{eq:y_ij_data}
\end{equation}
To detect the data symbols of $k$-th UE, a detector $f(\cdot)$ of choice can be applied onto the signal (\ref{eq:y_ij_data}), e.g. $\mathbf{\hat{d}}^{(k)}_{i,j} = f(\mathbf{t}^{(k)}_{i, j})$, where $\mathbf{t}^{(k)}_{i, j} = \frac{\mathbf{\tilde{y}}_{i, j}}{\sqrt{1 - \mathcal{E}^{(k)}_{i,j}}}$. For example, for \ac{LMMSE} detection, and approximating the interference as AWGN, $\mathbf{\hat{d}}^{(k)}_{i,j} =\left(\mathbf{G}_{i,j}^{(k)}\right)^{-1}\left(\mathbf{\hat{H}}^{(k)}_{i, j}\right)^{H}\mathbf{t}^{(k)}_{i, j}$, where $\mathbf{G}_{i,j}^{(k)} = \left(\mathbf{\hat{H}}^{(k)}_{i, j}\right)^{H} \mathbf{\hat{H}}^{(k)}_{i, j} + \sigma^{2} \mathbf{I},$ where $\sigma^{2}$ is the noise variance.
 
 A more advanced receiver may iterate between channel estimation, data detection, and interference cancellation as follows. At iteration $u$, the channel estimates for user $k$ are updated after the interference has been removed:
 \begin{align}
\mathbf{\hat{H}}_{i,j}^{(k, u)} &= \frac{\left( \mathbf{y}_{i,j} -\mathbf{v}^{(k, u-1)}_{i,j} \right) \left( \mathbf{p}_{i,j}^{(k)} \right)^H }{ \sqrt{\mathcal{E}^{(k)}_{i,j}} } \label{eq:LS_chan_k_closed_form-iter} \\
& \mathbf{v}^{(k, u-1)}_{i,j} =  \sum_{k'=1}^{K} \mathbf{\hat H}^{(k', u-1)}_{i, j}\sqrt{1-\mathcal{E}^{(k')}_{i,j}}\mathbf{\hat d}^{(k', u-1)}_{i,j} - \nonumber \\
&\sum_{k'=1, k'\neq k}^{K}\mathbf{\hat H}^{(k', u-1)}_{i, j}\sqrt{\mathcal{E}^{(k')}_{i,j}} {\mathbf{p}}^{(k')}_{i,j}.
\label{eq:LS_chan_k_closed_form-iter-interf}
\end{align}

Additionally, the estimated channel can be refined by utilizing a sliding window of dimensions $(W_f, W_t)$ as follows:
\begin{align}
    \mathbf{\hat{{H}}}_{i,j}^{(k, u)} &= \frac{\sum_{m=-W_f/2+1}^{W_f/2}\sum_{n=-W_t/2+1}^{W_t/2}  \mathbf{\hat{H}}_{i+m,j+n}^{(k, u)}}{W_f W_t}.
\label{eq:chan_smooth}
\end{align}

After the updated channel estimates are computed, the detection of the data of the user $k$ in iteration $u$ can be performed with a detector $f(\cdot)$, such as \ac{LMMSE}, after removing the contribution of SI DMRS and other users $k'\neq k$ from the received signal:

\begin{align}
\mathbf{\hat{d}}^{(k,u)}_{i,j}
&= f\Bigg(
       \frac{1}{\sqrt{1 - \mathcal{E}^{(k)}_{i,j}}}
       \Big[
          \mathbf{y}_{i,j}
          - \sum_{k'=1}^{K}
            \mathbf{\hat{H}}^{(k',u)}_{i,j}
            \sqrt{\mathcal{E}^{(k')}_{i,j}}
            \mathbf{p}^{(k')}_{i,j}
\label{eq:data_detect_iter} \\[2pt]
&\quad
          - \sum_{\substack{k'=1 \\ k'\neq k}}^{K}
            \mathbf{\hat{H}}^{(k',u)}_{i,j}
            \sqrt{1 - \mathcal{E}^{(k')}_{i,j}}
            \hat{\mathbf{d}}^{(k',u)}_{i,j}
       \Big] \Bigg). \nonumber
\end{align}

Lastly, the iterative algorithm is summarized in Algorithm \ref{alg:iterative_receiver}.

\begin{algorithm}[ht]
\caption{Iterative Receiver with \ac{SI} \ac{DMRS} Transmission}
\label{alg:iterative_receiver}
\begin{algorithmic}[1]
\REQUIRE Received signal $\mathbf{y}_{i,j}$, pilots $\{\mathbf{p}^{(k)}_{i,j}\}$, energies $\{\mathcal{E}^{(k)}_{i,j}\}$, noise variance $\sigma^2$, number of users $K$, iterations $U$

\vspace{0.5em}
\STATE Initialize $\hat{\mathbf{H}}^{(k,0)}_{i,j}$ using Eq.~\eqref{eq:LS_chan_k_closed_form} for all $k$
\STATE Initialize $\hat{\mathbf{d}}^{(k,0)}_{i,j} \gets \mathbf{0}$ for all $k$
\FOR{$u = 1$ to $U$}
  \FOR{each user $k = 1$ to $K$}
    \STATE Compute interference estimate $\mathbf{v}^{(k, u-1)}_{i,j}$ using Eq.~\eqref{eq:LS_chan_k_closed_form-iter-interf}
    \STATE Update $\hat{\mathbf{H}}^{(k,u)}_{i,j}$ using Eq.~\eqref{eq:LS_chan_k_closed_form-iter}
    \STATE Smooth $\hat{\mathbf{H}}^{(k,u)}_{i,j}$ using Eq.~\eqref{eq:chan_smooth}
    \STATE Compute detection input using Eq.~\eqref{eq:data_detect_iter}
    \STATE Estimate data $\hat{\mathbf{d}}^{(k,u)}_{i,j} = f(\cdot)$ (e.g., LMMSE as below Eq.~\eqref{eq:y_ij_data})
  \ENDFOR
\ENDFOR
\RETURN $\{\hat{\mathbf{d}}^{(k,U)}_{i,j}\}$ and $\{\hat{\mathbf{H}}^{(k,U)}_{i,j}\}$ for all $k$
\end{algorithmic}
\end{algorithm}
Note that channel decoder can be placed in or out of the loop of the iterative receiver.

\subsubsection{Architecture of DeepRx with SI DMRS transmission}
\label{sec:sip_deeprx}

Figure~\ref{fig:deeprx_block_diag} depicts the unified DeepRx architecture used both for \ac{SI} \ac{DMRS} as well as for orthogonal \ac{DMRS}. Focusing first on detection of \ac{SI} \ac{DMRS} transmissions, the input of the DeepRx model consists of the received \ac{SI} \ac{DMRS} signal $\mathbf{y}_{i, j}$, alongside with \emph{a priori} knowledge of the used \ac{SIP} configuration $\mathbf{p}_{i,j}^{(k)}$ for all $i$, $j$, and $k$. The received signal is processed first using the \ac{LS}-based  channel estimator, as described in Section~\ref{sec:classic_sip_chest}. This yields a channel estimate $\hat{\mathbf{H}}_{i,j}^{(k)}$, which is processed with a deep convolutional ResNet (Channel estimation CNN), whose input includes also the received signal and the transmitted superimposed pilots. Note that multiple versions of \ac{LS}-based  channel estimations, each with varying sliding window sizes, can be input to the Channel estimation CNN block. The output of the block, denoted by $\tilde{\mathbf{H}}_{i,j}^{(k)}$, can be interpreted as a refined channel estimate, which is used in the consecutive equalization part.

The equalization step within DeepRx consists of two parallel operations, one of which corresponds to \ac{LMMSE}-type equalization where a diagonal channel covariance matrix is assumed, while the other performs \ac{MRC}-type equalization. The outputs of these two parallel equalizer blocks are concatenated and fed to another deep convolutional ResNet. Conceptually, the first part of the deep ResNet (Detector CNN) can be considered to perform data-aided derotation and descaling of symbol estimates, while the latter part (Demapper CNN) maps the refined symbol estimates to log-likelihood ratios (LLRs). This interpretation stems from the fact that the detector CNN part is regularized with the transmitted symbols, which enforces the detector ResNet output to represent a type of soft symbol estimate. The refined soft symbol estimate is denoted by $\hat{\mathbf{d}}_{i,j}^{(k)}$ and Detector CNN also outputs additional hidden state to the Demapper CNN. Note that the Demapper CNN contains only 1x1 sized filters (corresponding to a single fully connected network applied for each RE separately), thus making it incapable of utilizing cross-RE information. The output of the demapper CNN part, consisting of LLRs and denoted by $\mathbf{l}_{i,j}^{(k)}$, represents the output of the DeepRx model and is fed to the LDPC decoder.

The complete DeepRx model, including also the non-learned channel estimator and equalizer blocks, can be trained end-to-end, assuming that we have a differentiable implementation of all the included blocks. In this work, the training is done similarly to \cite{Honkala21} by using the binary cross entropy (BCE) between the detected and transmitted bits as the loss function. With this, the BCE loss term can be written as follows:
\begin{align}
\mathrm{BCE} \left(\bm{\theta} \right) = -\frac{1}{W} \sum_{w=0}^{W - 1} & \left( b_{w} \operatorname{log}\left(\hat{b}_{w}\right) \right.\nonumber\\
&+ \left. \left(1-b_{w}\right) \operatorname{log}\left(1-\hat{b}_{w} \right) \right),
\end{align}
where $b_{w}$ is the transmitted bit, $\hat{b}_{w}$ is the bit probability estimate at the DeepRx output (after applying the Sigmoid function to convert LLRs to probabilities \cite{Honkala21}), and $W$ is the total number of transmitted bits within the slot. Additionally, the loss function includes a symbol-based regularization term.

\subsection{Orthogonal transmission}
In orthogonal \ac{DMRS} transmission, a portion of resource grid is reserved for \ac{DMRS} transmission. In \ac{5G} systems, the \ac{DMRS} of different transmission layers/users are designed to be orthogonal in time, frequency or code domain. Therefore, there is no intra- or inter-layer \ac{DP} interference and also the received \acp{DMRS} are almost orthogonal. allowing accurate channel estimation at the configured resources for \ac{DMRS} transmission. However, channel response should be estimated at the resources configured for data transmission. Therefore, an interpolation is needed to estimate the channel at data \acp{RE} based on the \ac{DMRS} \acp{RE}. Moreover, allocating resources specifically for \ac{DMRS} transmission decreases the system's achievable throughput, and the throughput degradation increases with the number of transmission layers/users.

\begin{table}[t]
\centering
\caption{Simulation Parameters}
\label{tab:simulaton_parameters}
\begin{tabular}{l l}
    \toprule
    \textbf{Parameter} & \textbf{Value} \\
    \midrule
    Number of users $K$ & $1, 2, 4$ \\
    Number of antennas at UE $N_T$ & $1, 2$ \\
    Number of antennas at BS $N_R$ & $4, 16$ \\
    Carrier frequency & $3.5$ GHz \\ 
    Subcarrier spacing & $30$ kHz \\ 
    FFT Size & $128$ \\
    Number of OFDM symbols & $14$ \\
    Number of subcarriers & $72$ \\
    MCS & $7, 14, 19$ \\
    Constellation Size $M$ & $4, 16, 64$ \\
    Channel model & UMa  \\
    User velocity & $[1, 10]$ m/s \\
    Channel implementation  & frequency domain \\
    \bottomrule
\end{tabular}
\end{table}
\subsubsection{Least-square channel estimation with Orthogonal DMRS transmission}
\label{sec:orthogonal_dmrs_chest}
After OFDM demodulation, the received signal at a \ac{DMRS} \ac{RE} configured for the $k$-th user can be expressed as
\begin{equation}
   \mathbf{y}_{i, j} = \mathbf{H}^{(k)}_{i, j}  \mathbf{p}^{(k)}_{i,j} + \mathbf{n}_{i,j}.
   \label{eq:y_ij_orth}
\end{equation}
An \ac{LS} channel estimator aims to determine the channel between the \ac{BS} and the $k$-th user at the $i$-th subcarrier and the $j$-th \ac{OFDM} symbol as 
\begin{equation}
 \hat{\mathbf{H}}_{i,j}^{(k)} = \arg\min_{\mathbf{H}_{i,j}^{(k)}} \left\| \mathbf{y}_{i,j} - \mathbf{H}_{i,j}^{(k)} {\mathbf{p}}^{(k)}_{i,j} \right\|_2^2,
   \label{eq:LS_chan_k_orth}
\end{equation} which results in: 
\begin{align}
\mathbf{\hat{H}}_{i,j}^{(k)} &= \mathbf{y}_{i,j} \left( \mathbf{p}_{i,j}^{(k)} \right)^H = \mathbf{H}^{(k)}_{i,j} \nonumber + \mathbf{n}_{i,j} \left( \mathbf{p}^{(k)}_{i,j} \right)^H.
\label{eq:LS_chan_k_closed_form}
\end{align}
It is important to highlight that we use a linear interpolation method followed by a frequency smoothing method to estimate the channel response at the data \acp{RE}.
\subsubsection{Architecture of DeepRx with Orthogonal DMRS transmission}

As mentioned in Section~\ref{sec:sip_deeprx}, the DeepRx model used for receiving signals with orthogonal \ac{DMRS} is essentially the same as for receiving \ac{SI} \ac{DMRS} transmissions. Referring to Fig.~\ref{fig:deeprx_block_diag}, the only difference is in the channel estimator block, which is done differently for orthogonal \ac{DMRS}. Now, the channel estimation is carried out as described in Section~\ref{sec:orthogonal_dmrs_chest}, and this estimate is fed to the ResNet block for further refining before the equalization blocks. The remainder of the processing within DeepRx is identical as when using it for the detection of \ac{SI} \ac{DMRS}  transmissions.

It should be noted, however, that in this work we trained separate DeepRx models for \ac{SI} \ac{DMRS} and for orthogonal \ac{DMRS}. Therefore, even though the model architectures are essentially the same for the two cases, the model weights are different.

\section{Performance evaluation}\label{section:Evaluation}

In order to gain insights on the performance of the different \ac{DMRS} and receiver schemes, we simulated a wide range of various configurations, both in SIMO and single-and multi-user MIMO with varying UE and antenna counts. In this study, we use Sionna for the link-level simulations \cite{sionna}.

\subsection{Simulation setup}

Table~\ref{tab:simulaton_parameters} provides information on the considered configuration parameters in the numerical results. Note that for all the scenarios, UMa \ac{3GPP} channel model is adopted, assuming an UE velocity range of 1--10 m/s. The utilized MIMO configurations range between 4 and 16 BS antennas and 1--4 spatial streams, i.e., MIMO layers. This ensures that the performance of the different schemes is evaluated under a wide range of scenarios. 

For the orthogonal DRMS transmission, we used DMRS port 0 for supporting one transmission stream, DMRS ports 0 and 2 for supporting two transmission streams and DMRS ports 0 to 3 for supporting four transmission streams.

For the \ac{SI} \ac{DMRS} transmission, we assume the same power ratio across all layers, UEs, and resource elements (REs). Consequently, we omit the indexing in $\mathcal{E}_{i,j}^{(k)}$ and refer to it simply as $\mathcal{E}$ for clarity.
\autoref{tab:receiver_parameters} presents the considered \ac{SI} \ac{DMRS} power ratio $\mathcal{E}$ and the dimensions $(W_f, W_t)$ of sliding windows utilized for channel estimation in both the iterative receiver and DeepRx during \ac{SI} \ac{DMRS} reception. 
Note that the number of iterations in the iterative receiver, denoted as $U$, corresponds to the number of windows listed in the table. For instance, the first entry with window sizes $[(12, 14), (6, 14)]$ indicates that the iterative receiver performs $U=2$ iterations, with the first iteration using a $(12, 14)$ sliding window and the second iteration employing a $(6, 14)$ sliding window. It is crucial to emphasize that to achieve optimal performance from the iterative receiver, we incorporated an LDPC decoder in the loop and conducted an extensive search for the optimal \ac{SI} \ac{DMRS} power ratio $\mathcal{E}$, number of iterations, and sliding window sizes for the iterative receiver. For other types of receivers, especially DeepRx, the LDPC channel decoder is utilized only once.

\begin{table*}[t]
\centering
\caption{Power ratios in SI DMRS transmission and window sizes utilized in the iterative receiver and DeepRx}
\label{tab:receiver_parameters}
\begin{tabular}{*7c}
\toprule
\multicolumn{3}{c}{\textbf{Scenario}} &  \multicolumn{2}{c}{\textbf{Power Ratio} $\mathcal{E}$} & \multicolumn{2}{c}{\textbf{Window Size $(W_f, W_t)$}}\\
\midrule
$\#$Users $K$ & MIMO & Constellation $M$ & Iterative Receiver   & DeepRx    & Iterative Receiver   & DeepRx\\
\midrule
$1$    &  $1\times4$ & $4$ & $0.14$  & $0.035$ & $[(12, 14), (6, 14)]$ & $[(12, 14), (6, 14)]$ \\
$1$    &  $1\times4$ & $16$ & $0.22$  & $0.035$ & $[(8, 14), (6, 14), (6, 14), (4, 14)]$ & $[(12, 14), (6, 14)]$ \\
$1$    &  $1\times4$ & $64$ & $0.3$  & $0.035$ & $[(8, 14), (6, 14), (6, 14), (4, 14), (2, 14]$ & $[(12, 14), (6, 14)]$ \\
$1$    &  $2\times4$ & $4$ & $0.22$  & $0.07$ & $[(12, 14), (6, 14)]$ & $[(12, 14), (6, 14)]$ \\
$1$    &  $2\times4$ & $16$ & $0.35$  & $0.07$ & $[(8, 14), (6, 14), (6, 14), (4, 14)]$ & $[(12, 14), (6, 14)]$ \\
$1$    &  $2\times4$ & $64$ & $0.43$  & $0.07$ & $[(8, 14), (6, 14), (6, 14), (4, 14), (2, 14]$ & $[(12, 14), (6, 14)]$ \\
$2$    &  $1\times4$ & $16$ & $0.35$  & $0.07$ & $[(8, 14), (6, 14), (6, 14), (4, 14)]$ & $[(12, 14), (6, 14)]$ \\
$2$    &  $1\times4$ & $64$ & $0.43$  & $0.07$ & $[(8, 14), (6, 14), (6, 14), (4, 14), (2, 14]$ & $[(12, 14), (6, 14)]$ \\
$4$    &  $1\times16$ & $4$ & $0.24$  & $0.14$ & $[(12, 14), (6, 14)]$ & $[(12, 14), (6, 14)]$ \\
$4$    &  $1\times16$ & $64$ & $0.55$  & $0.14$ & $[(8, 14), (6, 14), (6, 14), (4, 14), (2, 14]$ & $[(12, 14), (6, 14)]$ \\
\bottomrule
\end{tabular}
\end{table*}

\subsection{Antenna and UE configurations}

\subsubsection{SU-MIMO}

\begin{figure*}
  \centering
  \setlength{\tabcolsep}{-2.6ex}  
  \setkeys{Gin}{width=0.55\textwidth,trim=0 2cm 0 1.0cm, clip}  
  \begin{tabular}{@{}cc@{}}
    \includegraphics{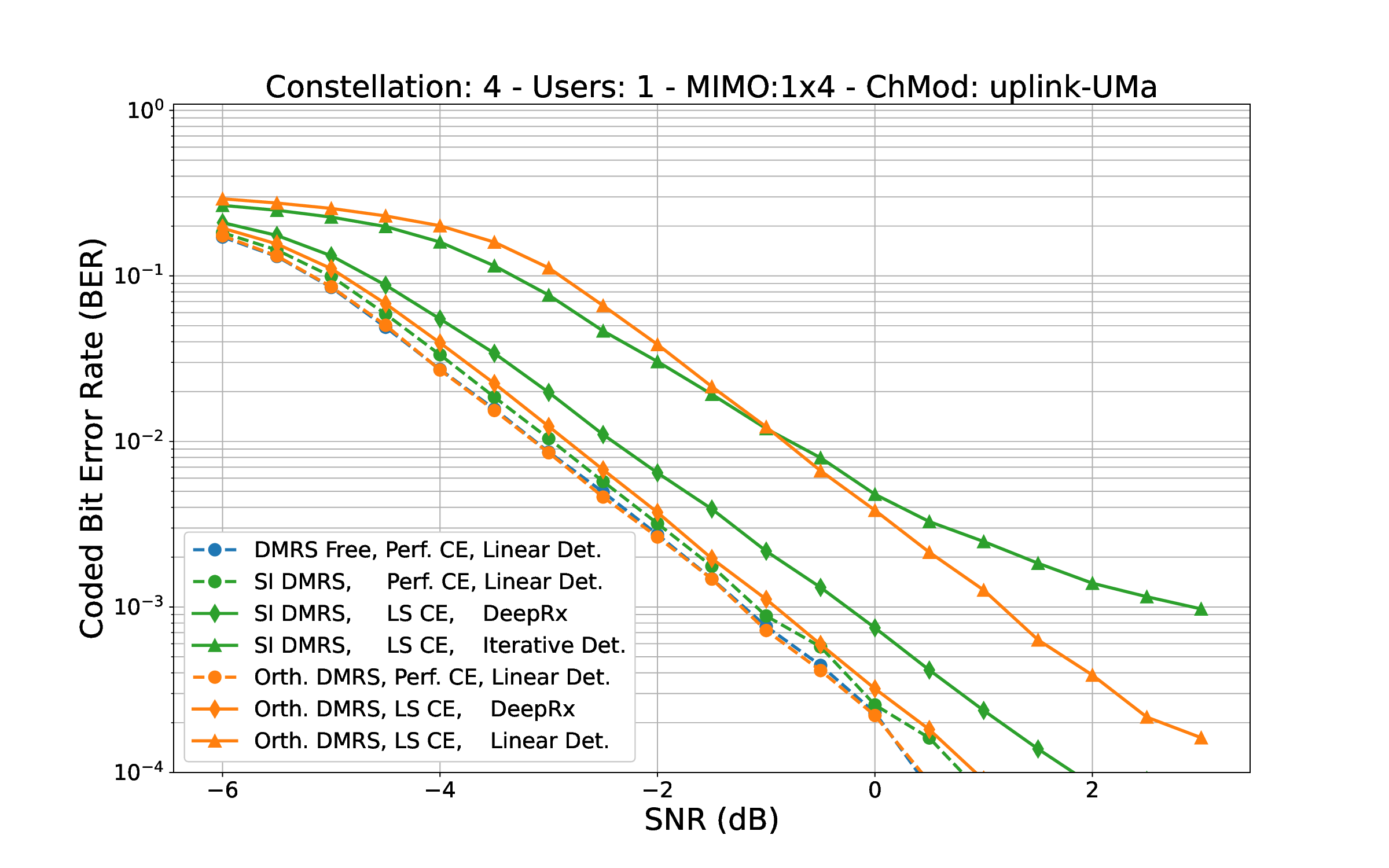} &
    \includegraphics{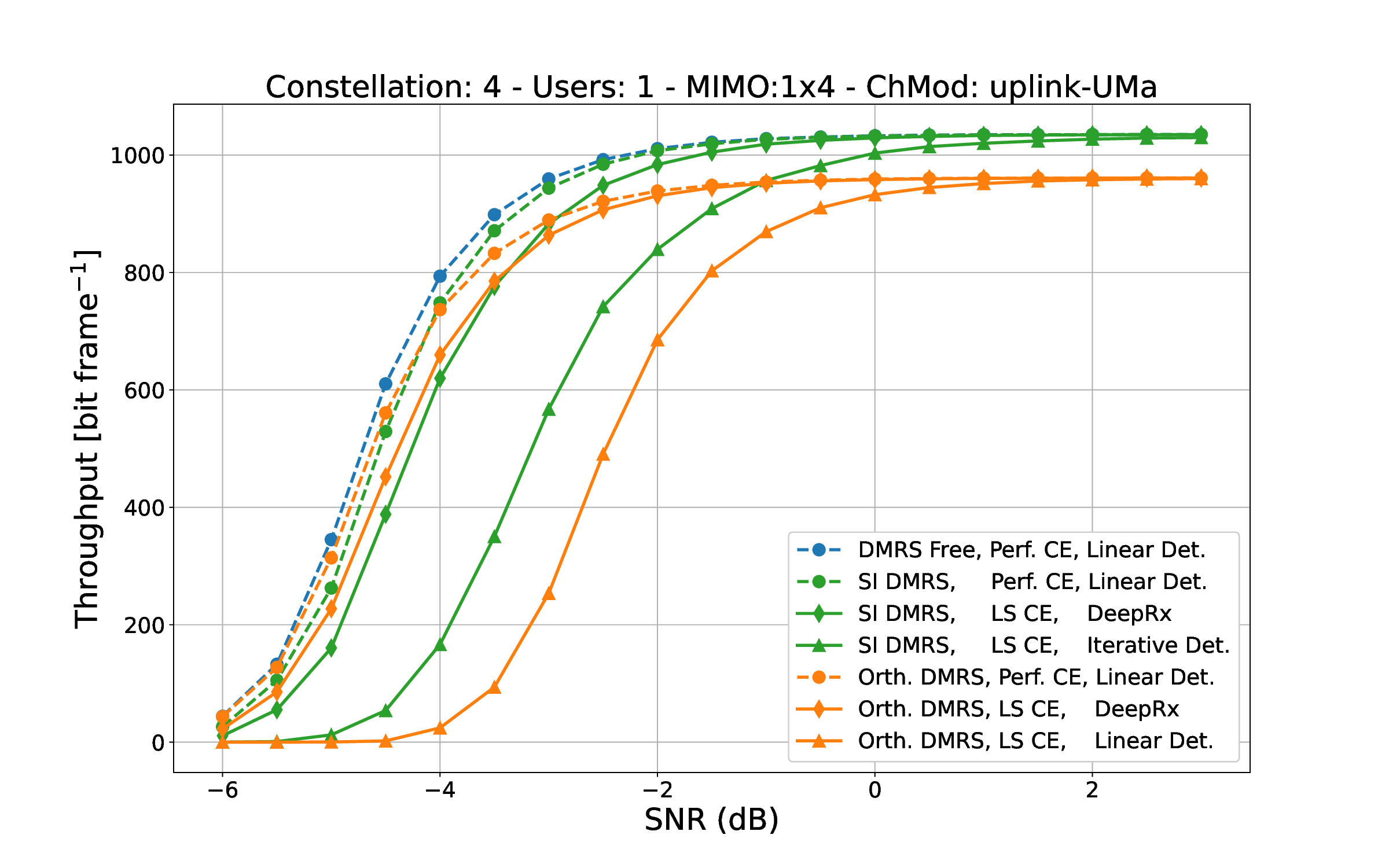} \\[3.0ex]
    \includegraphics{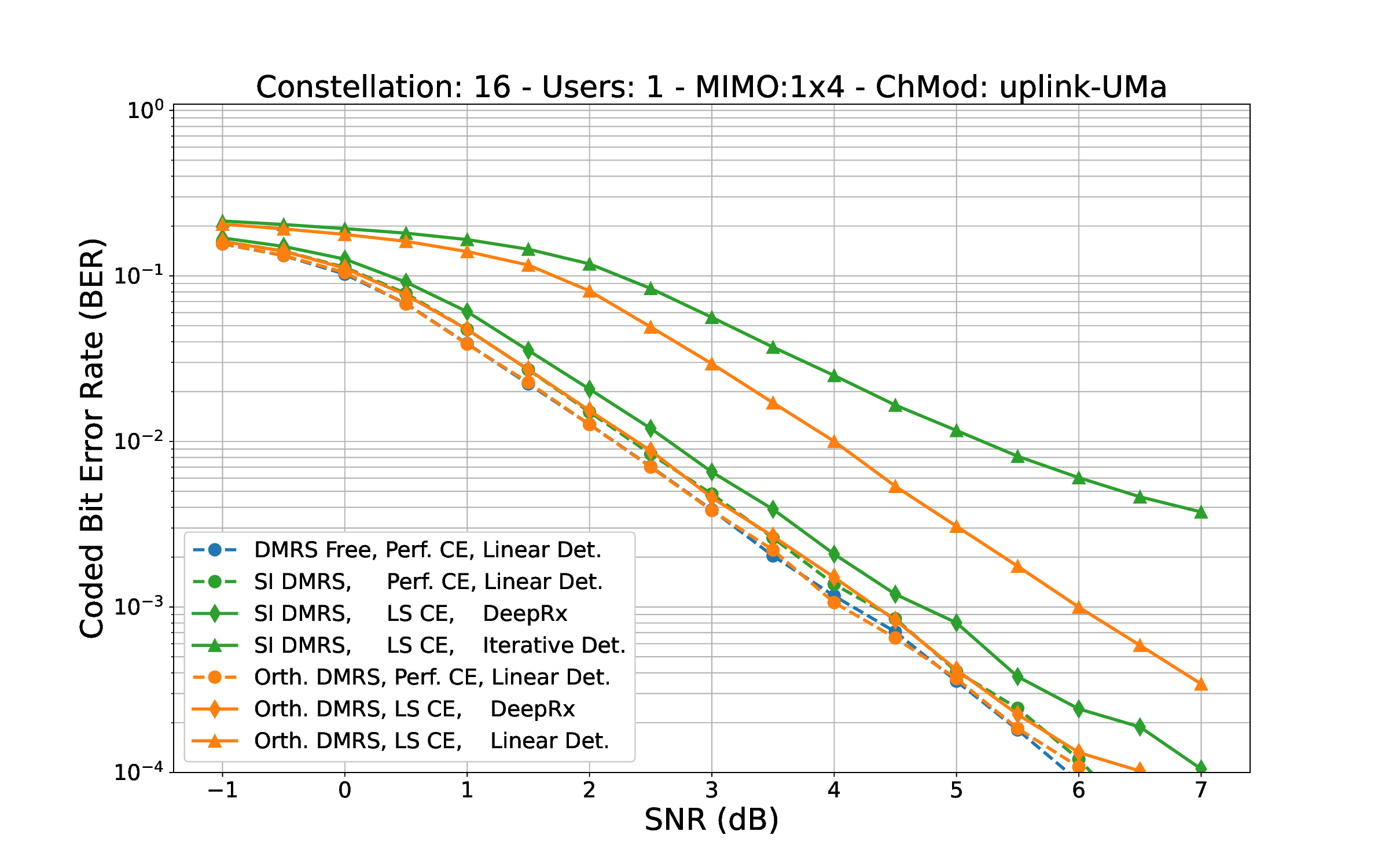} &
    \includegraphics{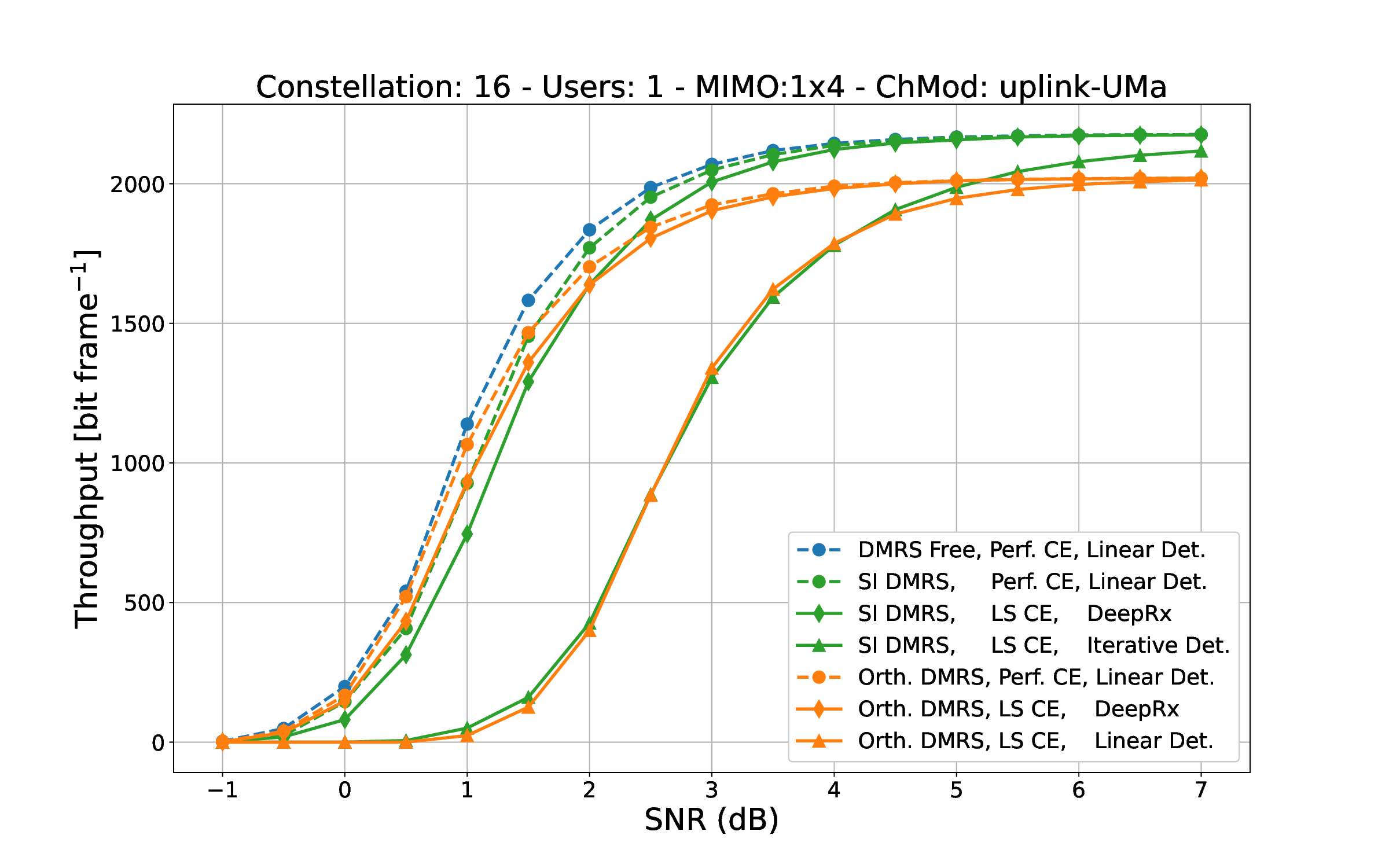} \\[3.0ex]
    \includegraphics{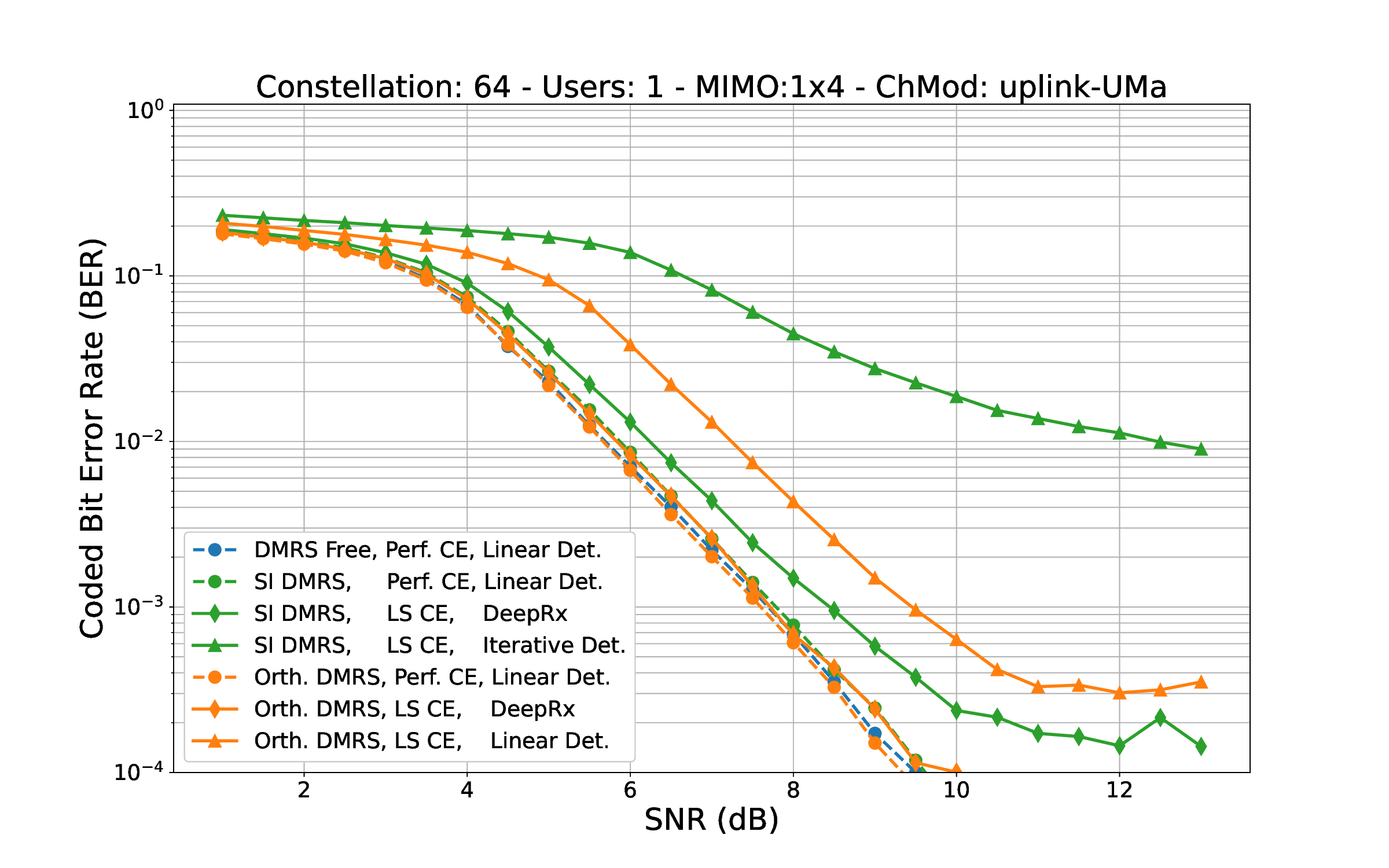} &
    \includegraphics{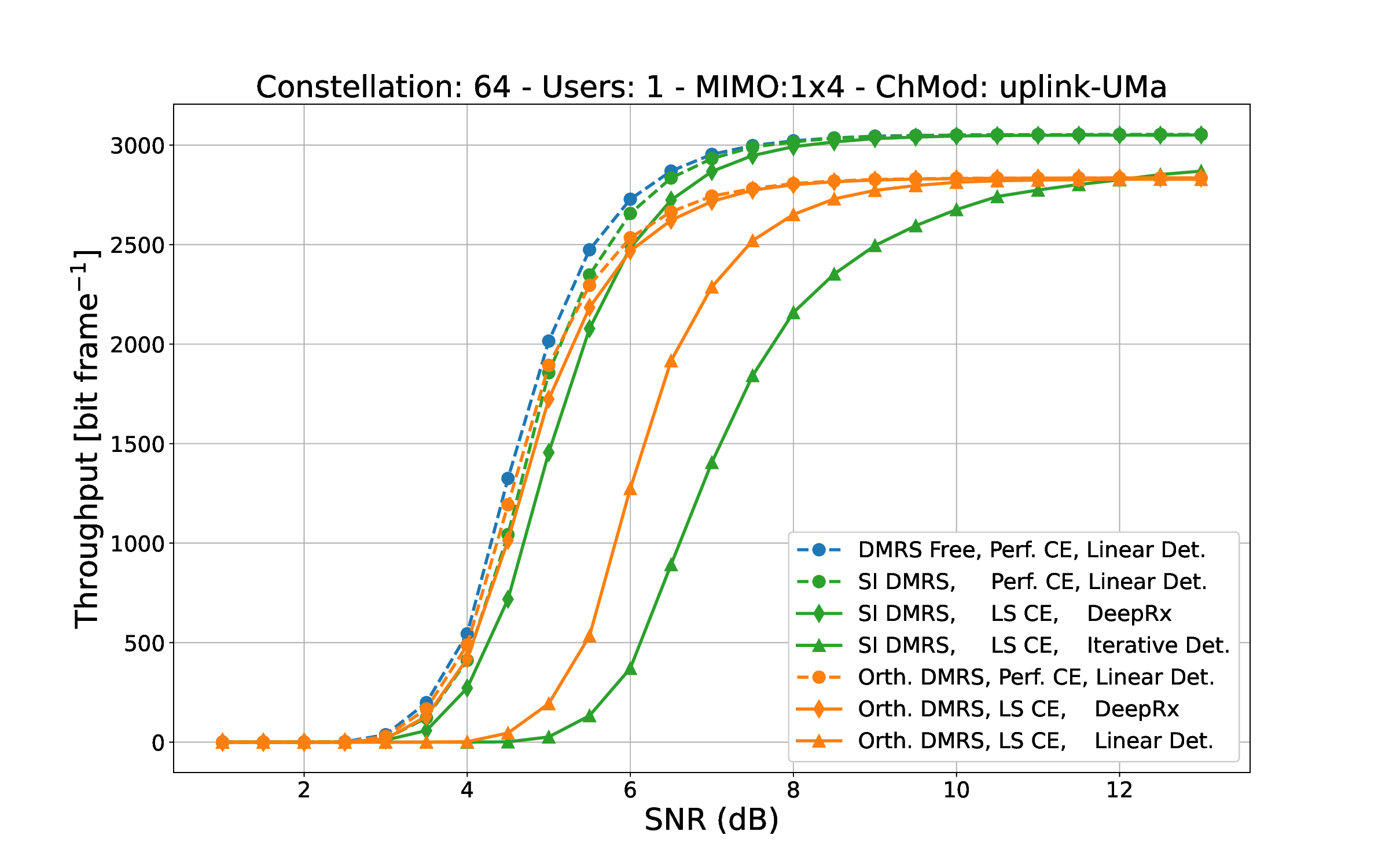}
  \end{tabular}
  \caption{Coded BER and throughput vs.\ SNR for SU MIMO 1x4 QPSK (top) 16QAM (middle) and 64QAM (bottom).}
  \label{fig:throughput_grid_su}
\end{figure*}

\begin{figure*}
  \centering
  \setlength{\tabcolsep}{-2.6ex}  
  \setkeys{Gin}{width=0.55\textwidth,trim=0 2cm 0 1.0cm, clip}  
  \begin{tabular}{@{}cc@{}}
    \includegraphics{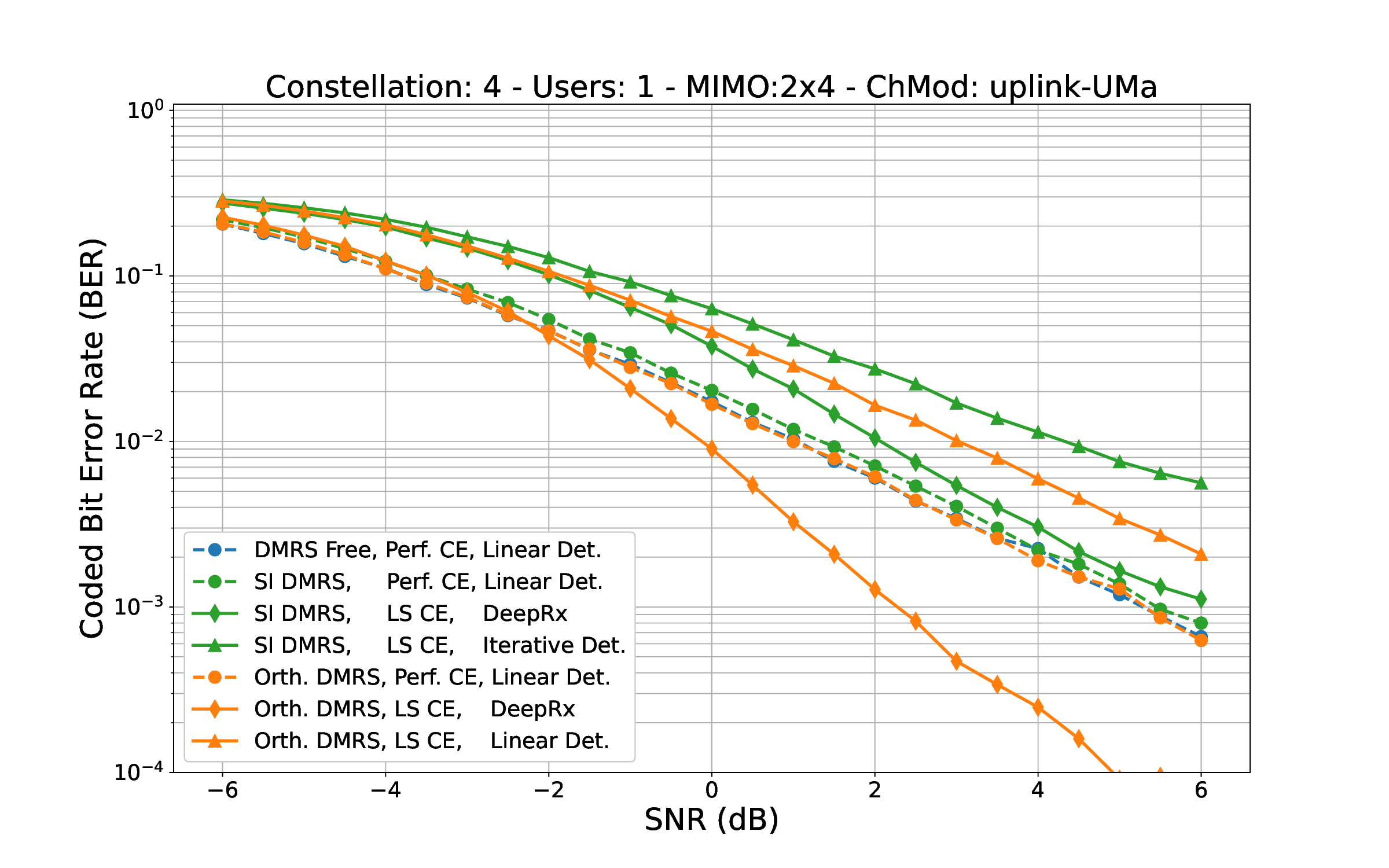} &
    \includegraphics{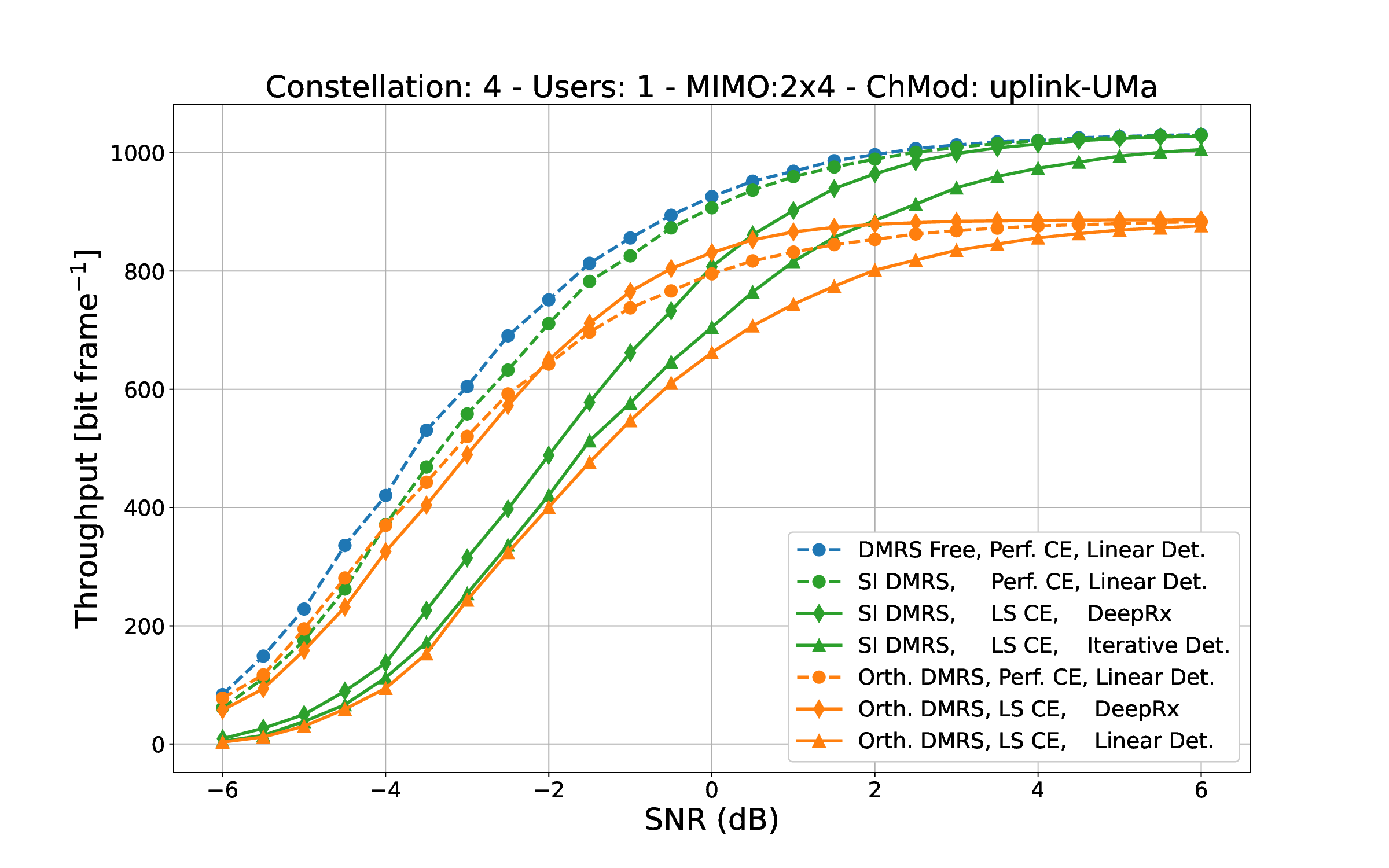} \\[3.0ex]
    \includegraphics{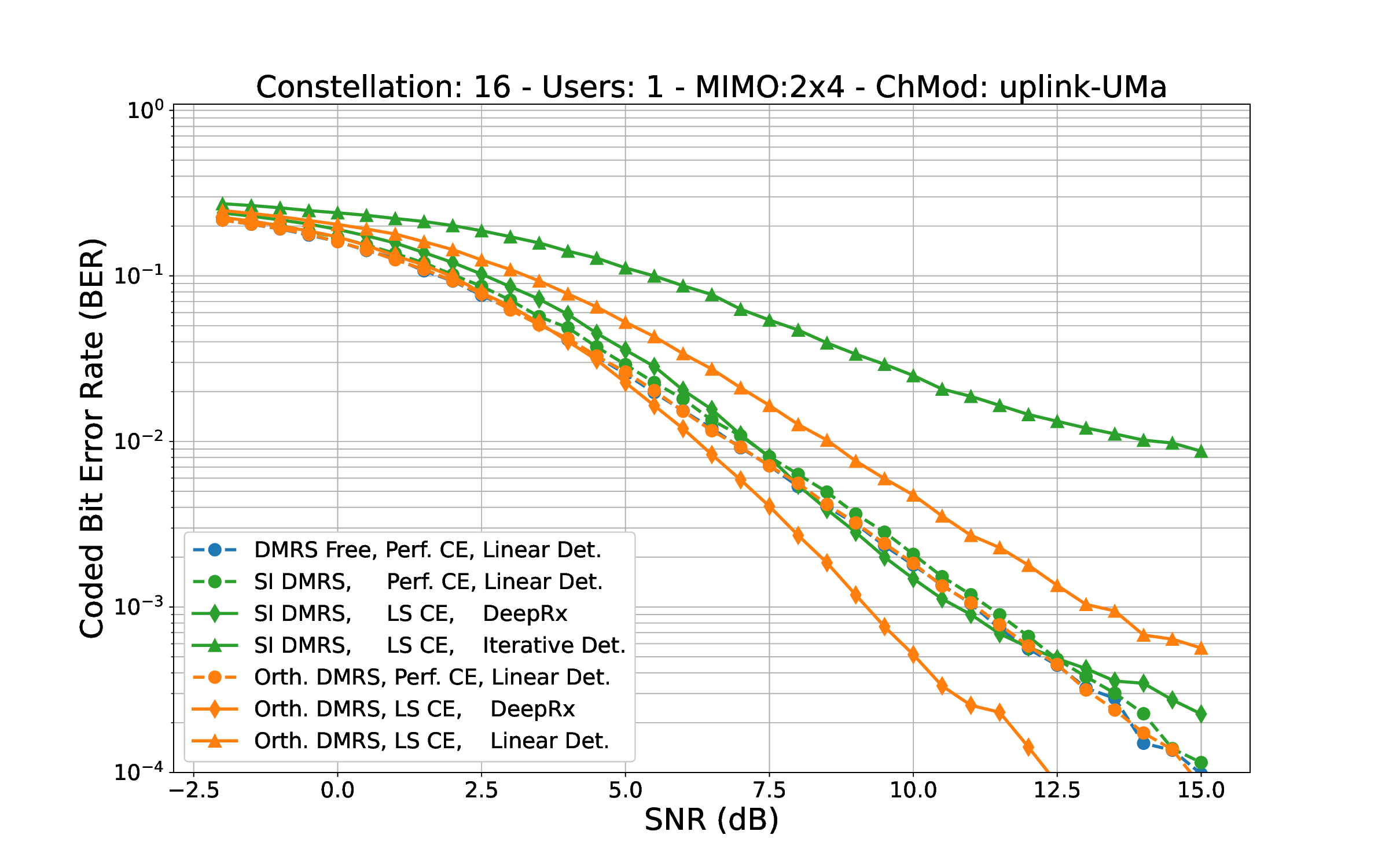} &
    \includegraphics{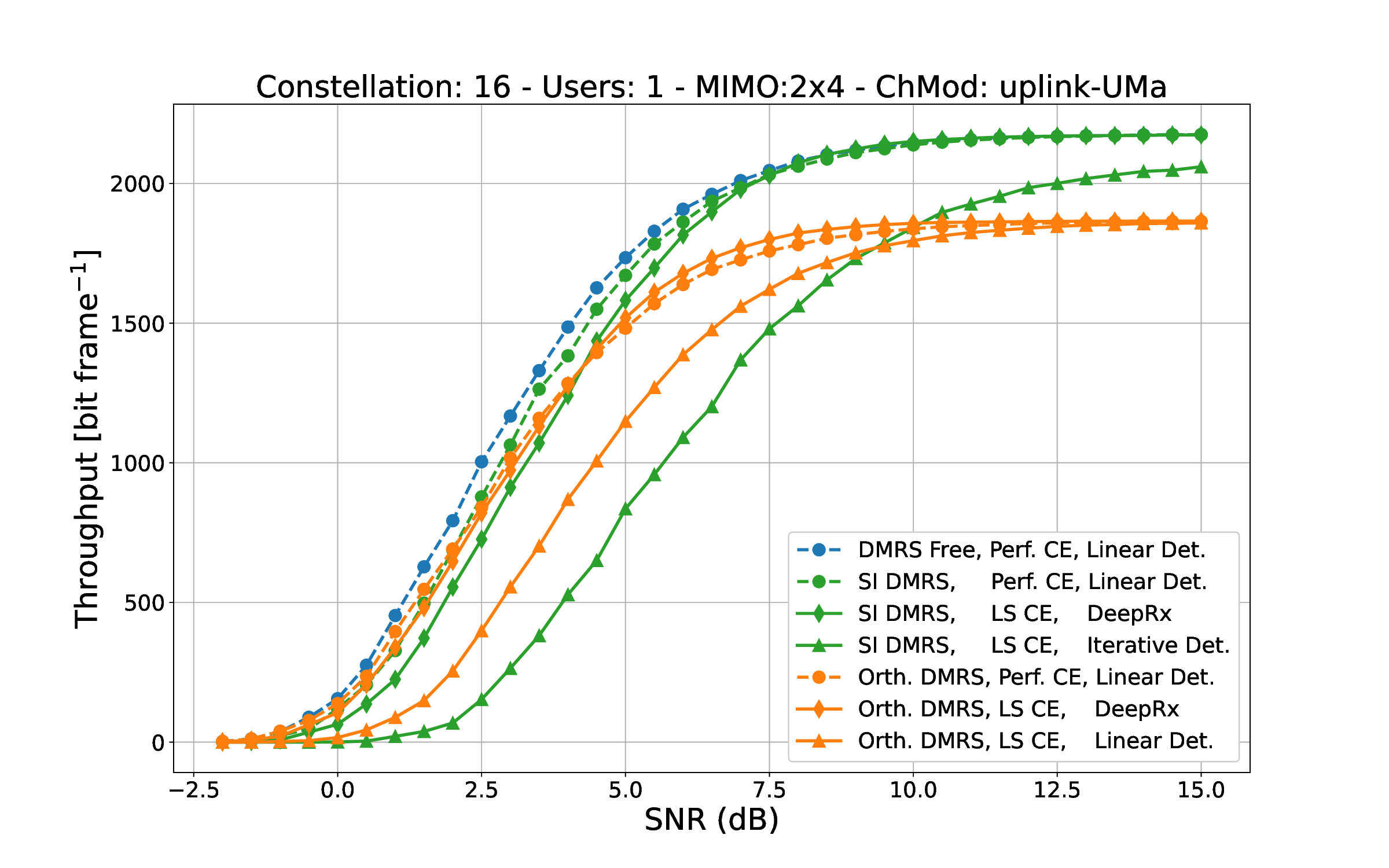} \\[3.0ex]
    \includegraphics{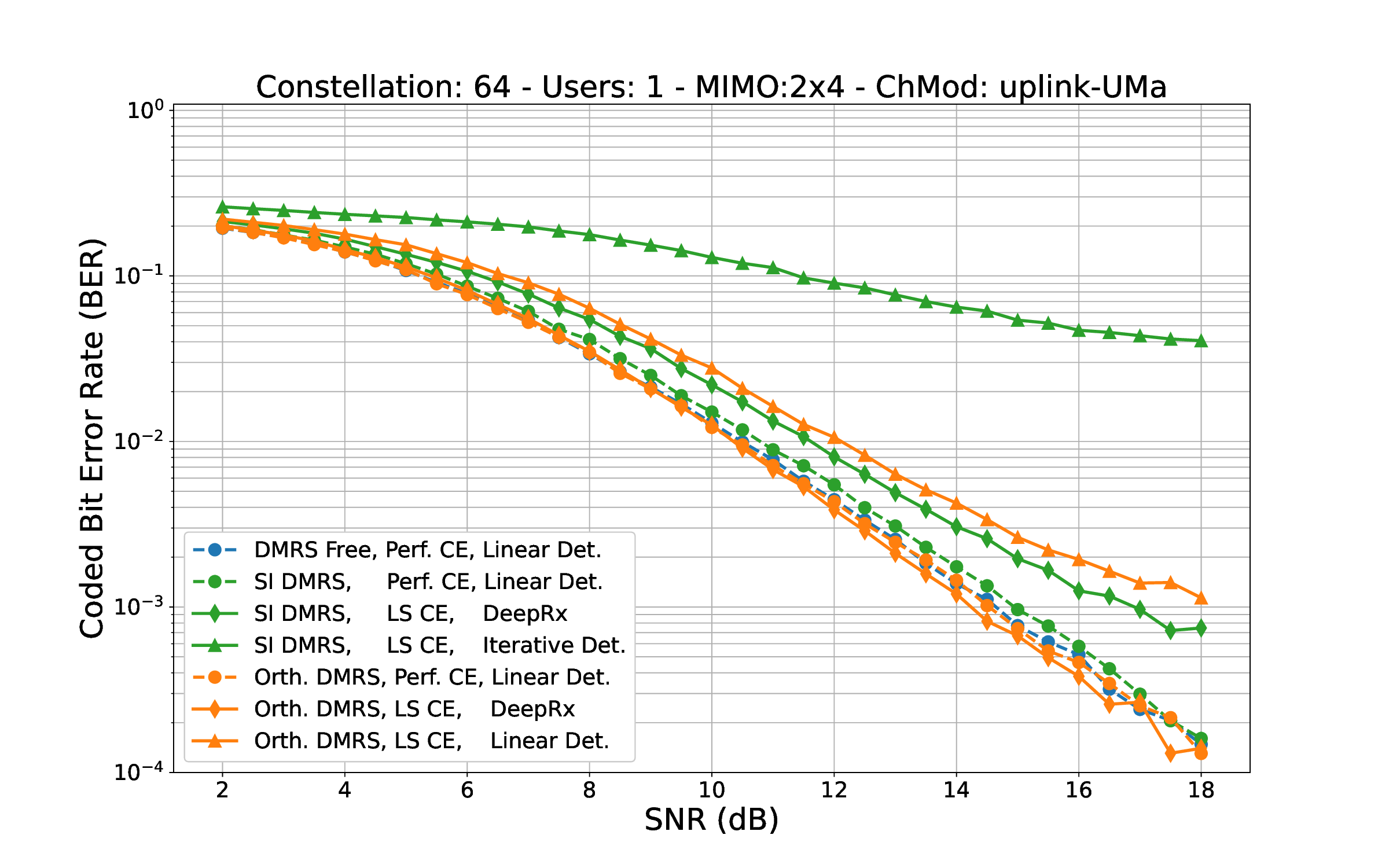} &
    \includegraphics{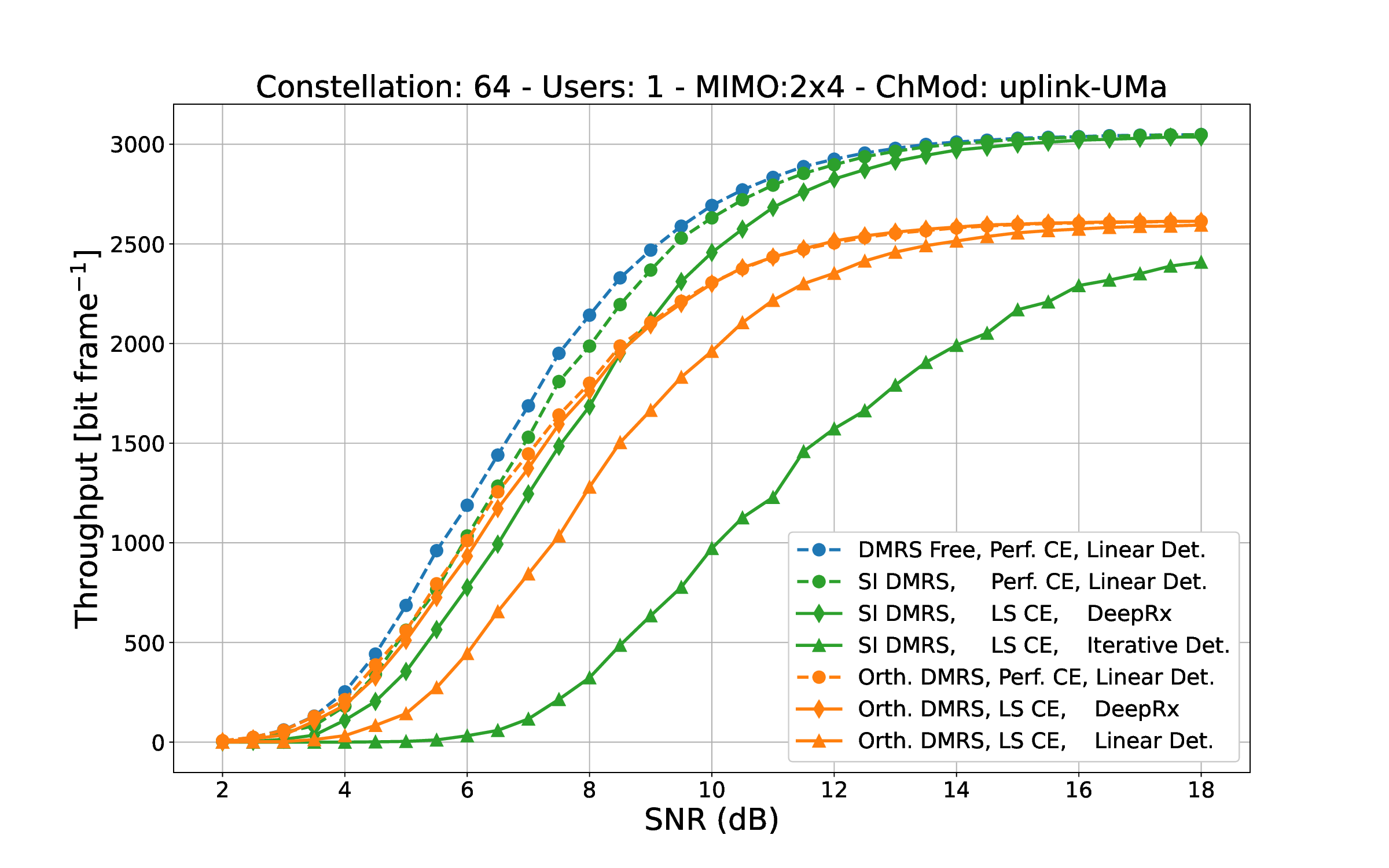}
  \end{tabular}
  \caption{Coded BER and throughput vs.\ SNR for SU MIMO 2x4 QPSK (top), 16QAM (middle), and 64QAM (bottom).}
  \label{fig:throughput_grid_su_2layer}
\end{figure*}

Let us first focus on the SU-MIMO case where the BS serves a single UE. The obtained throughputs of the different schemes under this scenario are shown for QPSK, 16-QAM, and 64QAM constellation and a single MIMO layer in Figure~\ref{fig:throughput_grid_su} and for two layers in Figure~\ref{fig:throughput_grid_su_2layer}. Note that the throughput is computed as a function of the \ac{BLER} using the expression 
$(1 - \text{BLER}) N_{d} \log_2(M)$, 
where $N_{d}$ represents the total number of allocated \acp{RE} used for data transmission.

Firstly, it can be observed that the ML-based schemes utilizing DeepRx outperform the conventional baselines in all of the six cases, both with \ac{SI} and orthogonal \ac{DMRS}. The performance gain of the DeepRx is especially pronounced in the single-layer case. With two layers, the gain is also evident, albeit not quite as large.

Another interesting aspect is the comparison between the \ac{SI} and orthogonal \ac{DMRS}. With a conventional receiver, the orthogonal \ac{DMRS} yields the better performance with 64-QAM modulation, while \ac{SI} DMRS is the better option under QPSK modulation. This is likely due to the higher sensitivity of the higher modulation order to any type of residual distortion due to the \ac{SI} \ac{DMRS}.

With the DeepRx-based schemes, the comparison between \ac{SI} and orthogonal \ac{DMRS} is more multi-faceted. In all cases, orthogonal \ac{DMRS} is preferable at low SNR, while \ac{SI} provides the higher throughput with high SNR. However, typically the difference at low SNR is rather small, while the gain from the \ac{SIP} becomes substantial at high SNRs. This is due to the fact that it has a higher upper limit for the spectral efficiency under good conditions, due to the fact that there are no dedicated resources for \ac{DMRS}. It should be noted that in a practical system, the schemes with orthogonal and \ac{SI} \ac{DMRS} might switch to a higher or lower modulation and coding scheme (MCS), which might vary the gain of the \ac{SIP} solution. Finally, it can be observed that with 2 layers and QPSK modulation, the performance of \ac{SI} DMRS at low SNR is quite far behind the scheme with orthogonal \ac{DMRS}. A possible explanation for this could be a random fluctuation in the training performance, or sub-optimal hyperparameterization, e.g. \ac{SI} \ac{DMRS} power ratio, for this particular scenario. In addition, it is important to note that in Orthogonal DMRS transmission, DeepRx provides higher throughput compared to the linear receiver with perfect channel knowledge. This is because the considered genie-aided receiver is a linear receiver, whereas DeepRx processing may be oriented towards non-linear receivers, such as the Maximum Likelihood receiver. Consequently, if a Maximum Likelihood receiver with perfect channel knowledge were considered, it would represent the upper-bound achievable throughput with orthogonal DMRS transmission.

\subsubsection{MU-MIMO}

\begin{figure*}
  \centering
  \setlength{\tabcolsep}{-2.6ex}  
  \setkeys{Gin}{width=0.55\textwidth,trim=0 2cm 0 1.0cm, clip}  

  \begin{tabular}{@{}cc@{}}
    \includegraphics{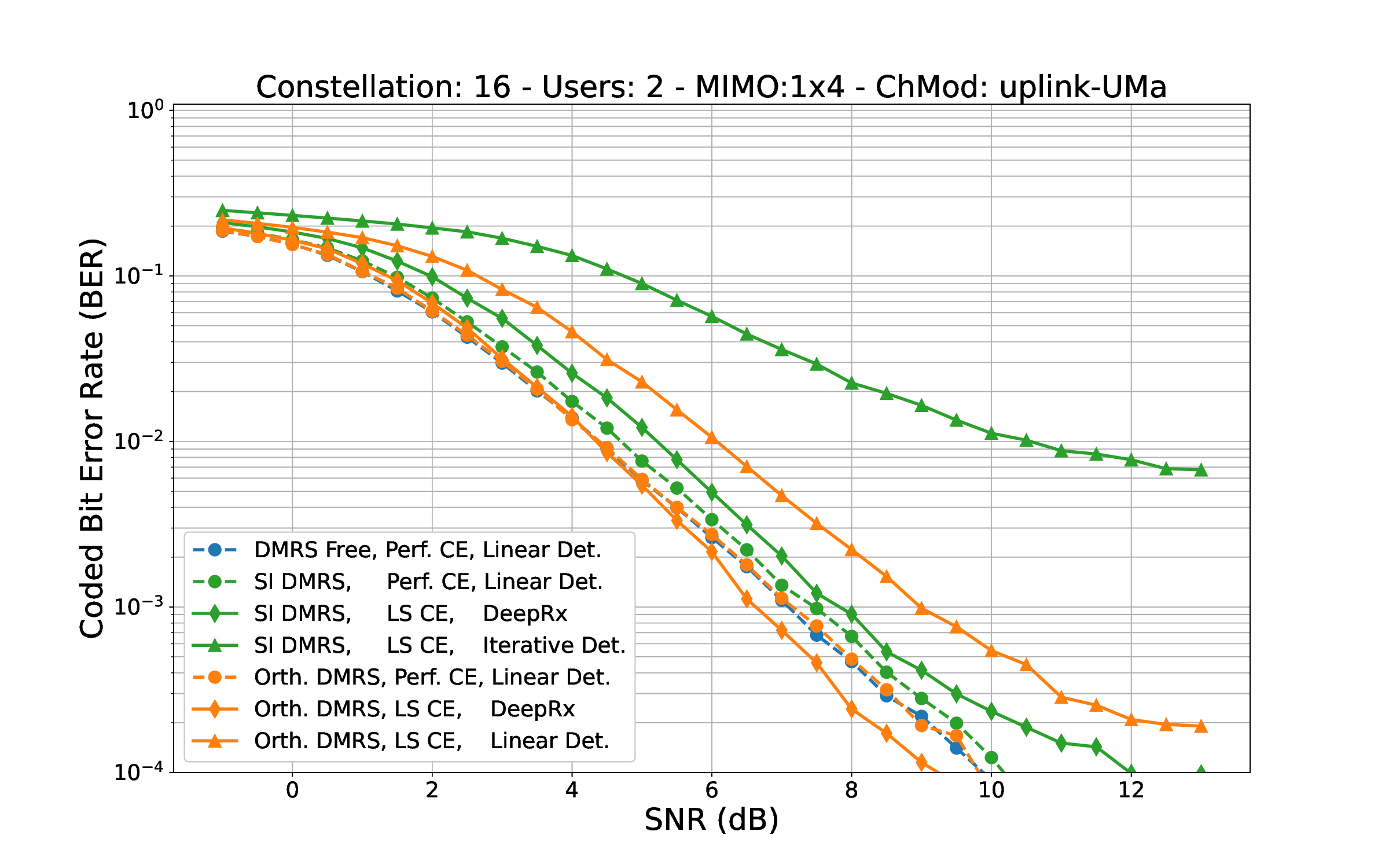} &
    \includegraphics{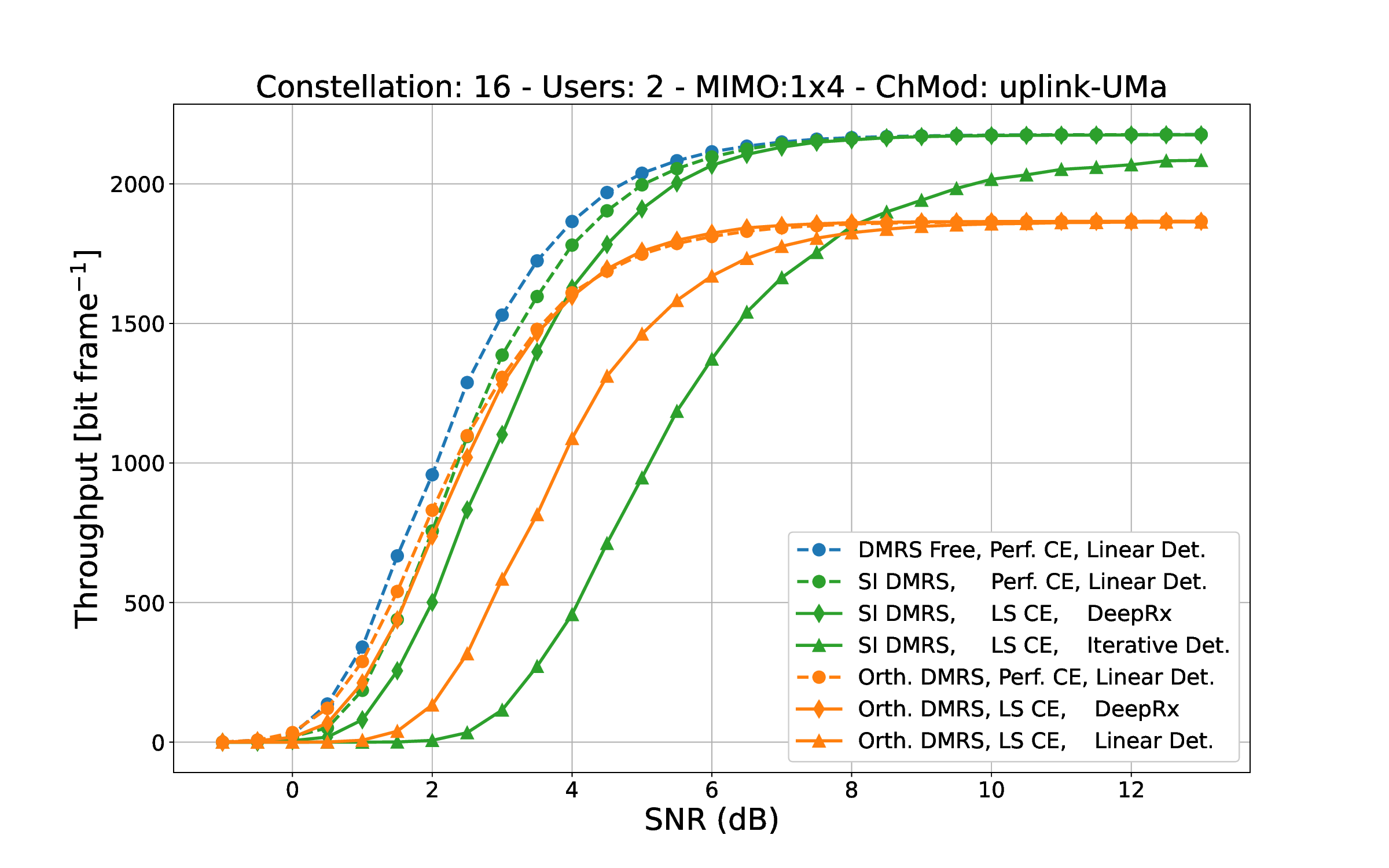} \\[2.1ex]
    \includegraphics{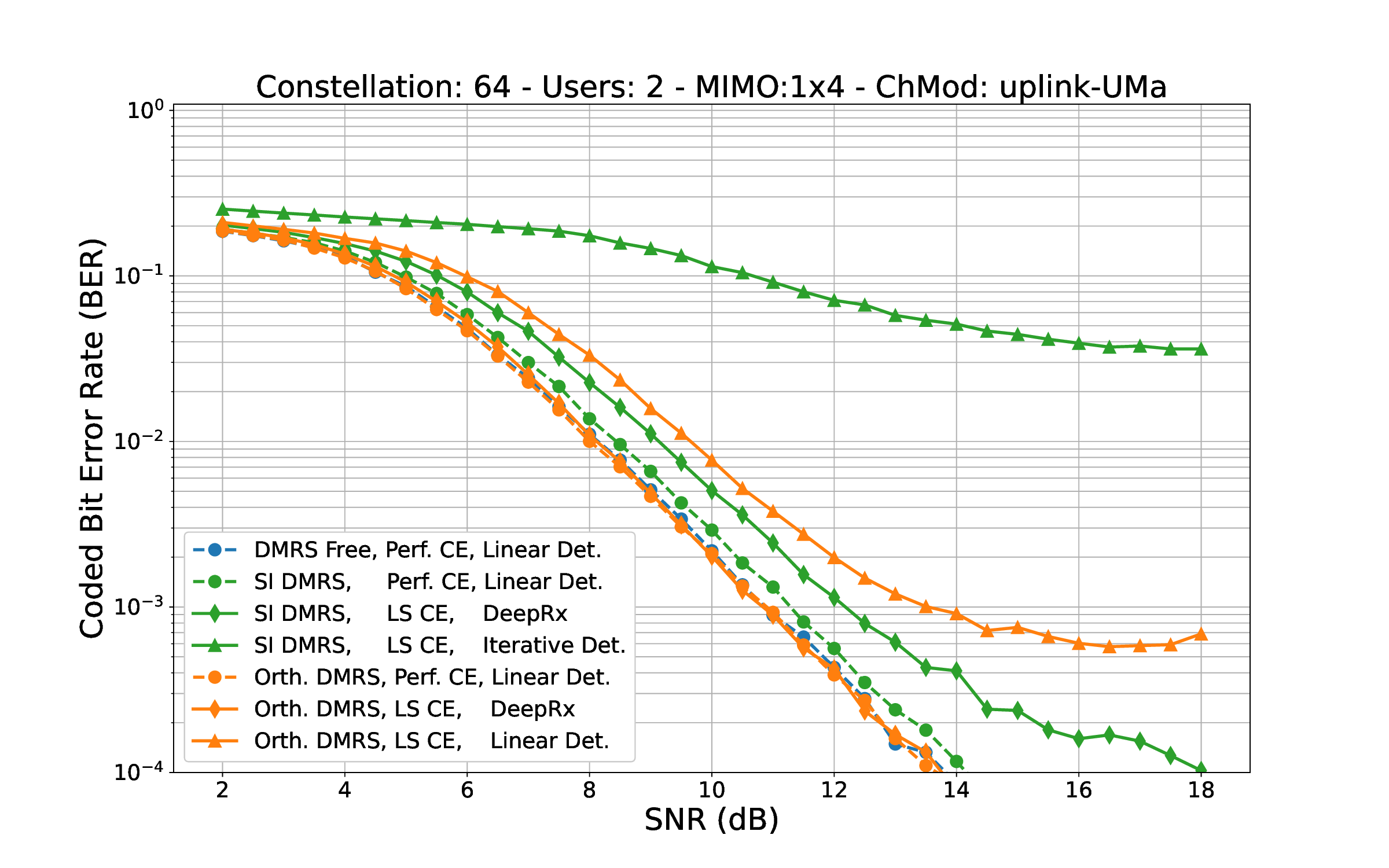} &
    \includegraphics{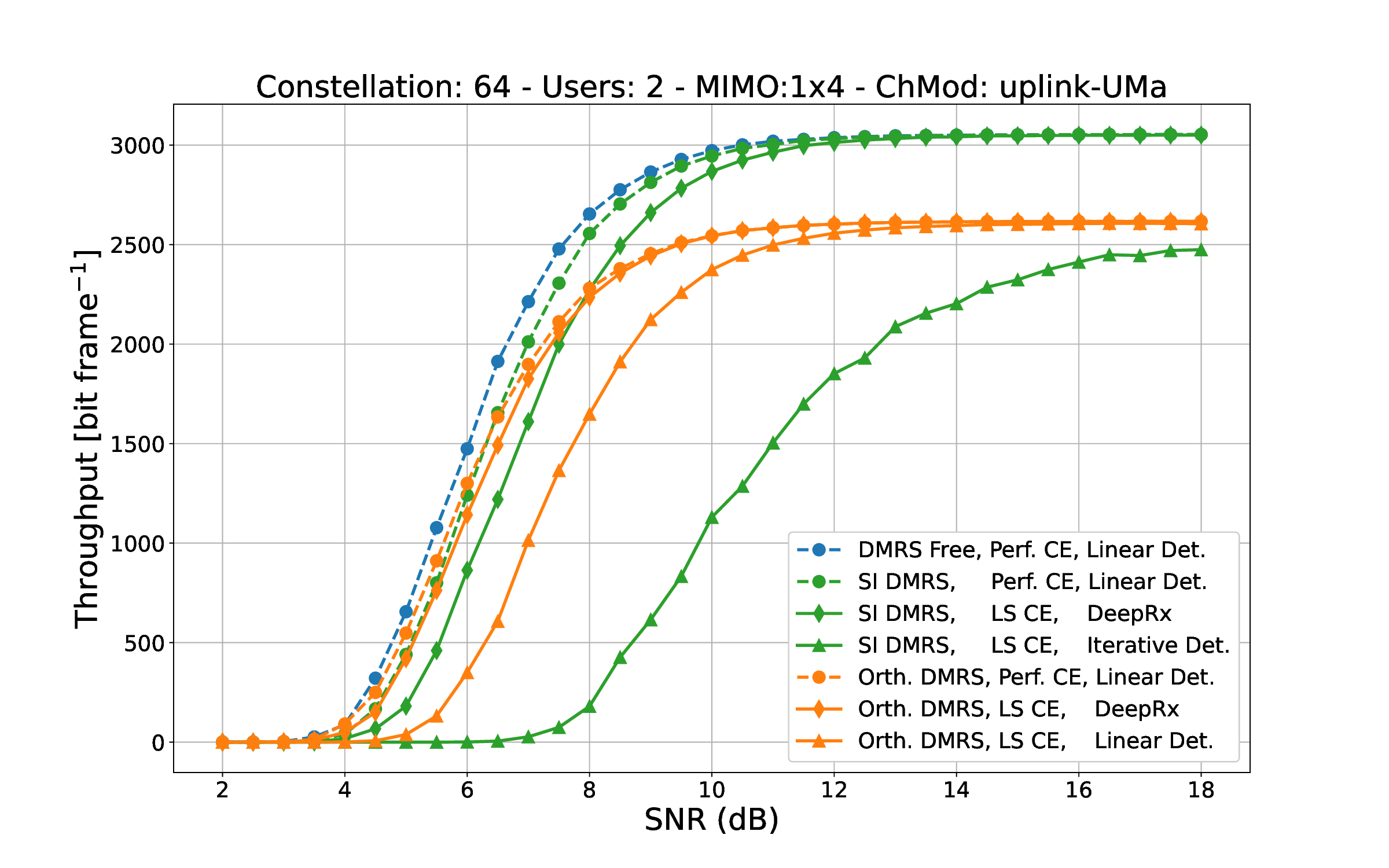}
  \end{tabular}
  \caption{Coded BER and throughput vs.\ SNR for MU MIMO, 2x(1x4) 16QAM (top) and 64QAM (bottom).}
  \label{fig:throughput_grid_mu}
\end{figure*}

\begin{figure*}
  \centering
  \setlength{\tabcolsep}{-2.6ex}  
  \setkeys{Gin}{width=0.55\textwidth,trim=0 2cm 0 1.0cm, clip}  

  \begin{tabular}{@{}cc@{}}
    \includegraphics{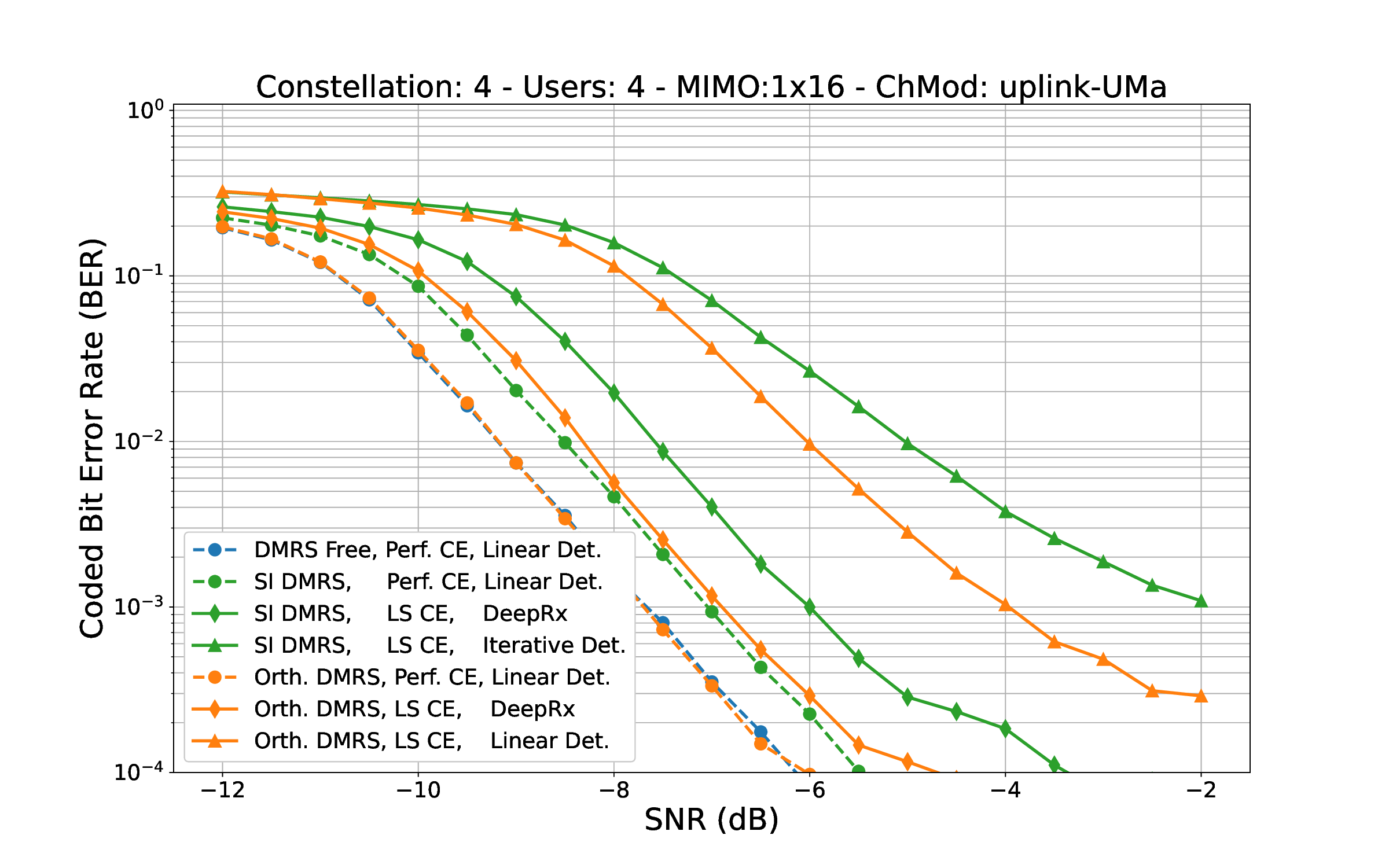} &
    \includegraphics{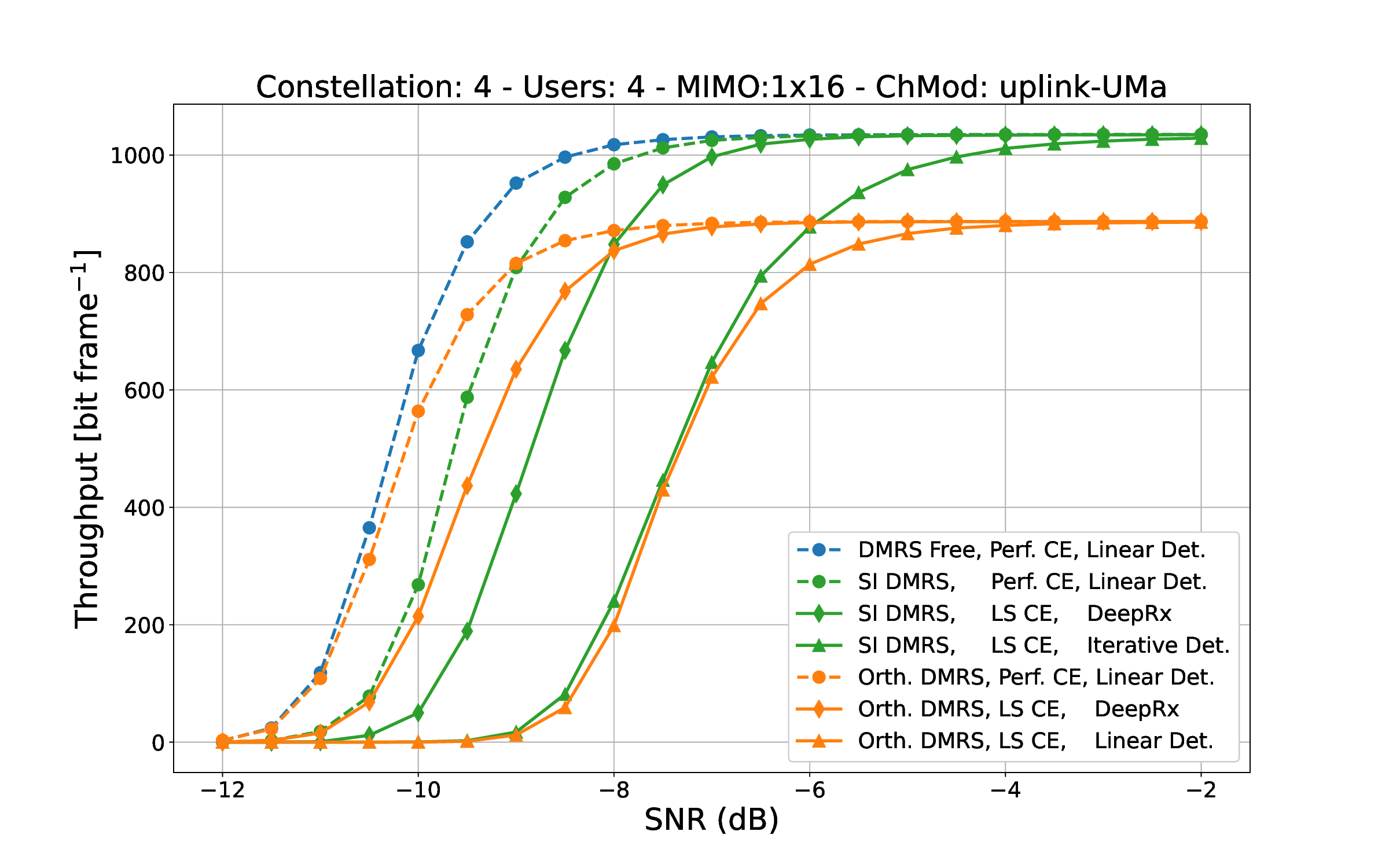} \\[2.1ex]
    \includegraphics{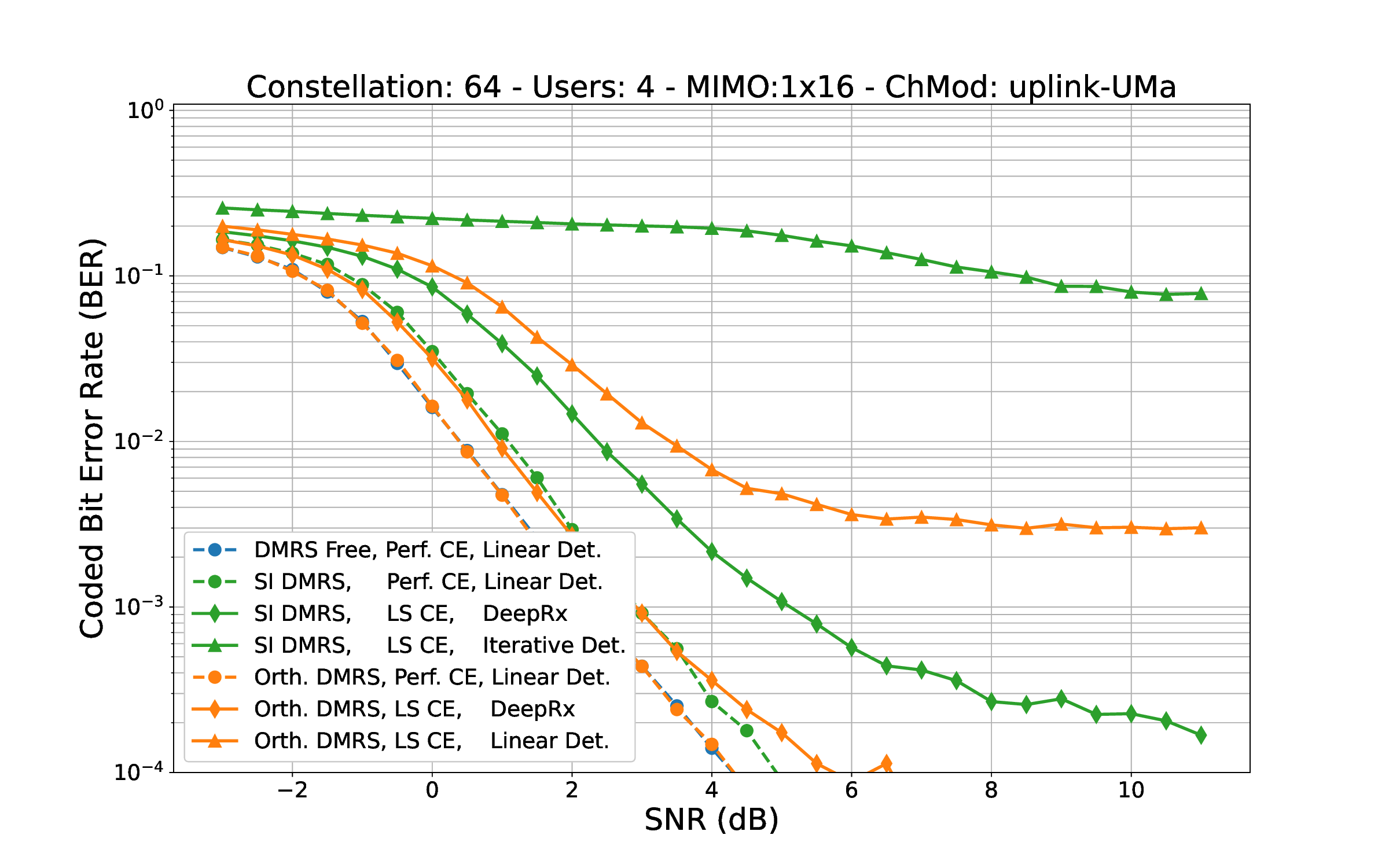} &
    \includegraphics{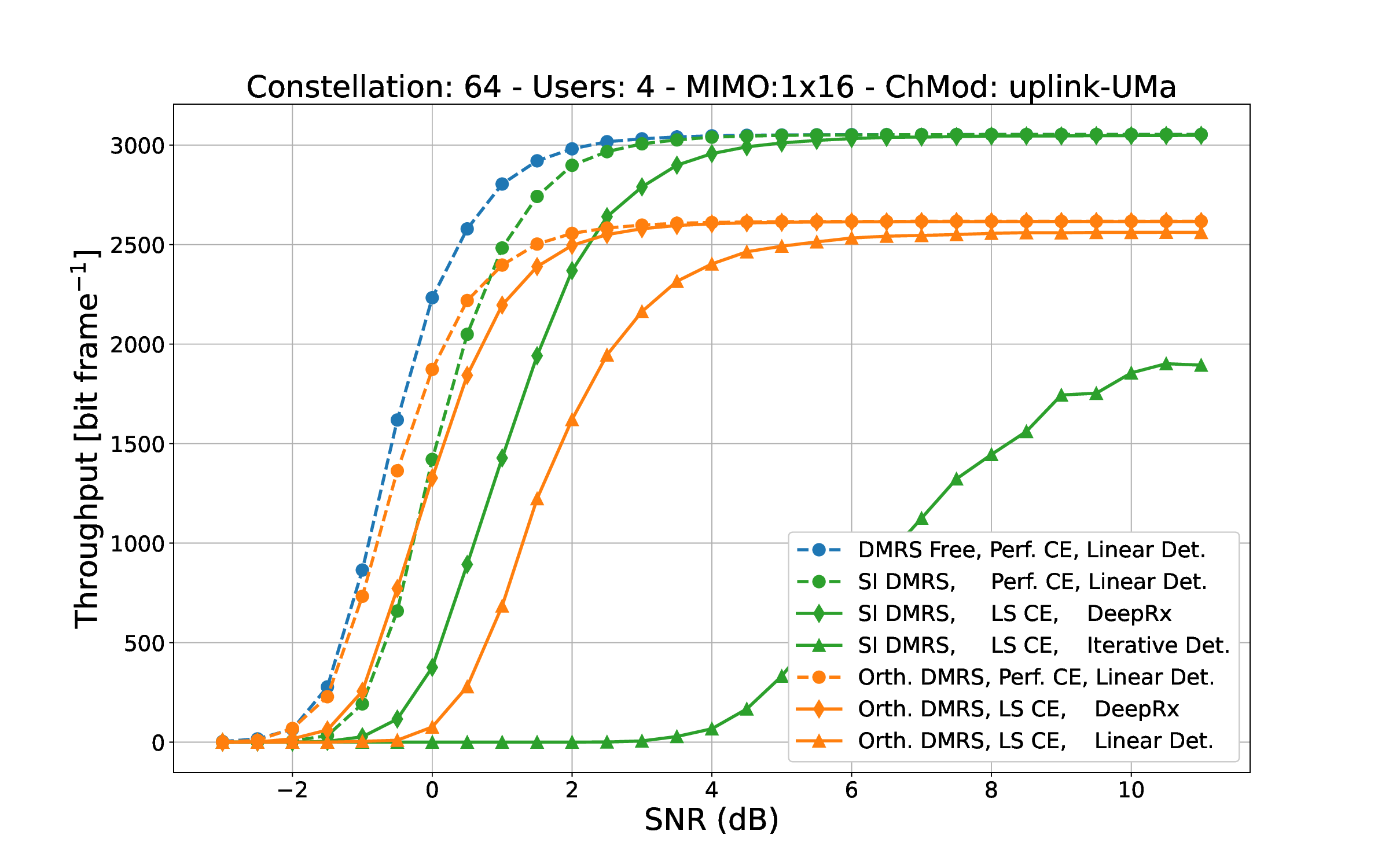}
  \end{tabular}
  \caption{Coded BER and throughput vs.\ SNR for MU MIMO, 4x(1x16) QPSK (top) and 64QAM (bottom).}
  \label{fig:throughput_grid_mu_4layers}
\end{figure*}

Next, let us evaluate the schemes under MU-MIMO scenario, where the BS employs 4 and 16 antennas to serve 2 or 4 users, emplying either QPSK, 16-QAM or 64-QAM constellation. The throughput results are collected in Figure~\ref{fig:throughput_grid_mu} for 2 UEs and in Figure~\ref{fig:throughput_grid_mu_4layers} for 4 UEs.

Regarding the conventional baseline, the conclusions in MU-MIMO scenario are largely similar to the SU-MIMO case. That is, it falls short of the throughput provided by the ML-based DeepRx schemes, especially with \ac{SI} DMRS. The only scenario where the conventional system achieves higher performance with \ac{SI} DMRS is the case with QPSK and 4 UEs. Again, this is most likely due to the high robustness of the QPSK modulation, as opposed to the more sensitive higher order modulations.

As for the DeepRx-based schemes, it is again observed that orthogonal DMRS yields the better performance at low SNR, while \ac{SI} DMRS outperforms it at high SNR. This is especially pronounced with 4 UEs, where the \ac{SI} DMRS scheme with DeepRx has somewhat lower performance at lower SNRs compared to orthogonal DMRS. This is most likely due to the rather challenging nature of detecting 4 spatial streams using only \ac{SI} DMRSs. With 2 UEs, the difference between \ac{SI} and orthogonal DMRS is rather small, when utilizing DeepRx.

Altogether, in the tested scenarios, employing DeepRx for reception suggests that \ac{SI} DMRS is likely to deliver enhanced performance across a wide SNR range for lower spatial multiplexing, and in higher SNR ranges for higher MIMO orders.

When looking at coded \ac{BER} of the proposed DeepRx-based and classical receivers for single-user and multi-user scenarios with 64-QAM modulation, in orthogonal \ac{DMRS} scheme, the DeepRx shows excellent performance, which is close to the performance of linear detector with perfect channel knowledge. However, in  \ac{SI} \ac{DMRS}, there is up to $1$dB gap between the performance of DeepRx and the genie-aided baseline assuming perfect channel knowledge. On the other hand, as shown in \autoref{fig:throughput_grid_su} and \autoref{fig:throughput_grid_mu}, the possibility of data transmission over all the \acp{RE} eventually results in throughput gain compared to Orthogonal \ac{DMRS}. In addition, the iterative receiver cannot provide good performance with \ac{SI} \ac{DMRS} in high modulation orders, due to lack of precise channel estimation.

\subsection{Neural Network Depth and Other Hyperparameters}

All CNNs consist of ResNet blocks:

\begin{equation}
	A_{i+1} = A_i + f\comp A_i,
	\label{eq:resblock}
\end{equation}

where the Demapper CNN uses only
$1\times1$ convolutions and ReLU nonlinearities,
and the other CNNs use

\begin{equation}
	f = 
	\mathrm{DW}_{9\times1}\comp
	\mathrm{ReLU}\comp
	\mathrm{DW}_{1\times9}\comp
	\mathrm{ReLU}
	\label{eq:fsubnet}
\end{equation}

i.e., two depthwise-separable 2D convolutions with asymmetric $1\times9$ and $9\times1$ kernels. We also alternate dilations of 1, 4, and 8 in frequency direction in adjacent ResNet blocks to increase the receptive field inside the network. For time direction, there is no dilations, since our frame has 14 symbols only. Compared to standard ResNet CNN architectures, this design achieves similar or better accuracy while reducing both complexity and parameter count.

For simplicity, the main results use the same network depth and hyperparameters across all scenarios:
12 ResNet blocks in both the Channel estimate CNN and the Detector CNN and 4 blocks in Demapper CNN (denoted \emph{Depth 12+12}). The only variable is the layer width (number of channels); we scale it in terms of $N_R$ and $N_T^{(k)}$ as specified in Table \ref{tab:hyperparameters}. We trained for 100k iterations using the Adam optimizer, a batch size of 128 and an initial learning rate of 0.002, which was reduced to zero according to a linear decay schedule by the end of training. Samples were generated on-the-fly and never reused, effectively rendering the dataset infinite.

\begin{figure}[!t]
\centering
\includegraphics[width=0.98\columnwidth]{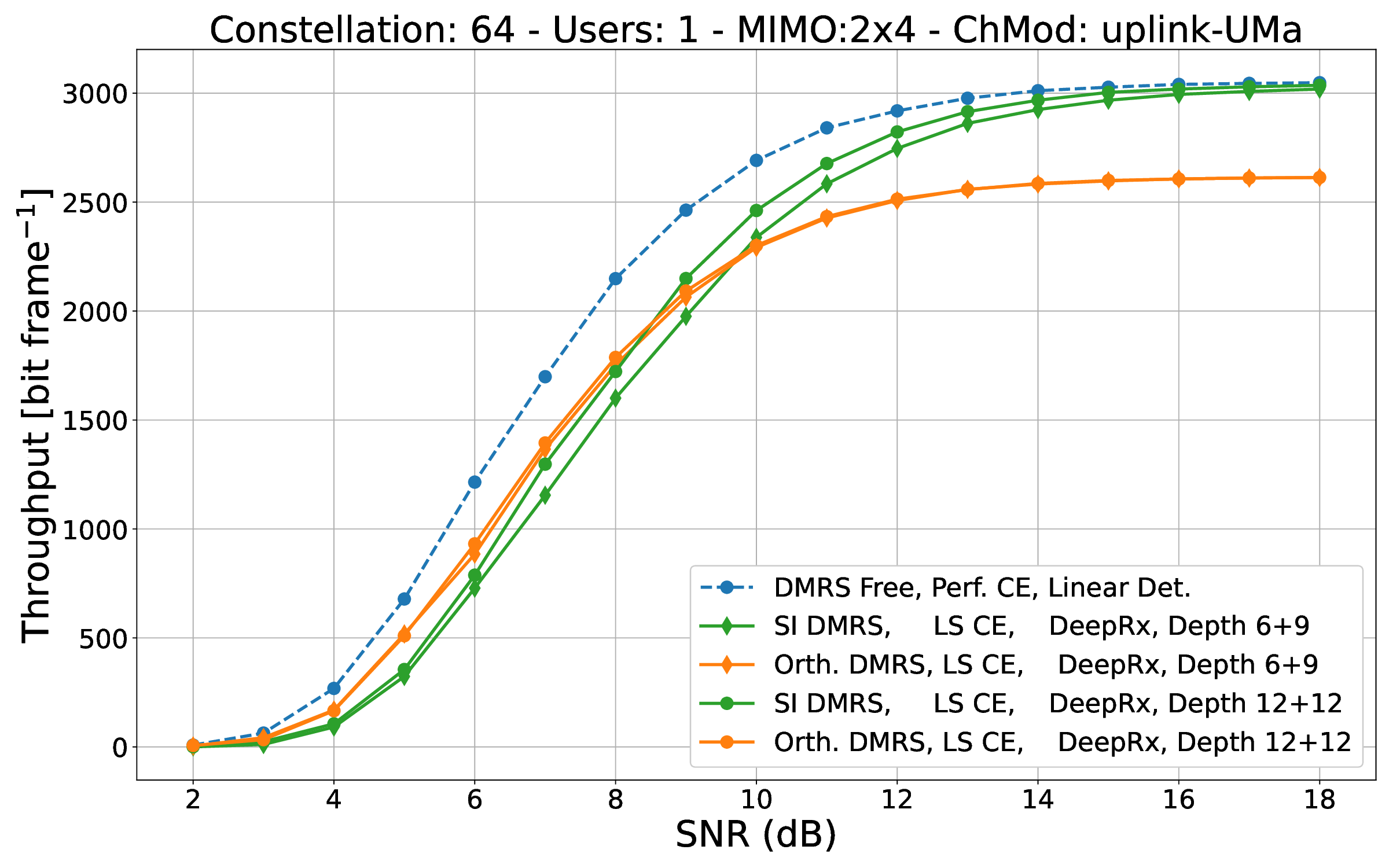}
\caption{Throughput results with different depths (number of ResNet blocks) of DeepRx.}
\label{fig:deeprx_depths}
\end{figure}

To study the effect of depth, we also evaluated a shallower network— \emph{Depth 6+9}—on the $4\times2$ SU MIMO with 64-QAM configuration (see Table \ref{tab:hyperparameters}). Note that we kept the Demapper CNN to static 4 blocks. 
Figure~\ref{fig:deeprx_depths} shows the achievable throughputs with different DeepRx model depths, considering both orthogonal and \ac{SI} DMRS. It can be observed that the shallower models with 6 and 9 ResNet sections achieve nearly comparable performance to the deeper model variant with 12 and 12 ResNet sections. The difference is somewhat larger with \ac{SI} DMRSs, indicating that it is a more challenging task for the DeepRx model. With orthogonal \ac{DMRS}, there is hardly any benefit from increasing the model size. In general there seems to be a graceful degradation as depth is reduced, but since this work does not target a specific
hardware platform, optimizing the hyperparameters for
concrete implementations is left for future work.

\begin{table}[ht]
  \centering
  \caption{Architectural settings for the three CNN modules}
  \label{tab:hyperparameters}
  \scalebox{0.86}{
  \begin{tabular}{*5c}
    \toprule
    \textbf{CNN part} & \textbf{Blocks 12+12} & \textbf{Blocks 6+9} & \textbf{Width formula}   & \textbf{Width 4×2} \\
    \midrule
    Chan.est.  & 12 & 6        & $16 \times N_R \times \sum_k N_T^{(k)} $  & 128   \\
    Detector   & 12 & 9        & $128 \times \sum_k N_T^{(k)}$             & 256    \\
    Demapper   & 4  & 4        & $64 \times \sum_k N_T^{(k)}$              & 128      \\
    \bottomrule
  \end{tabular}}
\end{table}

\section{Conclusions}\label{section:Conclusions}
In this work, we have conducted a comprehensive comparison between superimposed and orthogonal DMRS schemes across a range of SIMO and MIMO uplink configurations. We introduced a novel variant of DeepRx, a convolutional neural network-based receiver architecture, capable of performing joint channel estimation and data detection under superimposed DMRS. By overlaying pilot and data symbols, the superimposed DMRS approach enables more efficient utilization of time-frequency resources without compromising estimation accuracy.

Extensive evaluations in both single-user and multi-user uplink scenarios demonstrate that the proposed DeepRx variant outperforms conventional receivers in terms of throughput and maintains robust performance even as the number of users and transmission layers increases. These results suggest that DeepRx can effectively address the scalability challenges and overhead limitations inherent to orthogonal DMRS schemes, thereby enhancing spectral efficiency while keeping inference complexity within practical bounds.

Moreover, when a portion of the resource grid is reserved for control channels—as is the case with the first two OFDM symbols in 5G—the incentive to adopt superimposed DMRS becomes even stronger, as it allows more efficient use of the remaining data-carrying resources.

Future directions will focus on extending the proposed framework to more complex deployment scenarios, including large-scale antenna arrays and environments with significant inter-cell interference. We also plan to integrate advanced precoding and combining strategies, as well as explore mechanisms for online adaptation to dynamic channel conditions. 
Furthermore, optimizing the data-to-pilot power allocation and co-designing the superimposed reference signal structure jointly with the neural receiver architecture will be investigated to unlock additional performance gains.
\bibliographystyle{IEEEtran}

\bibliography{sp_journal}

\begin{IEEEbiography}[{\includegraphics[width=1in,height=1.25in,clip,keepaspectratio]{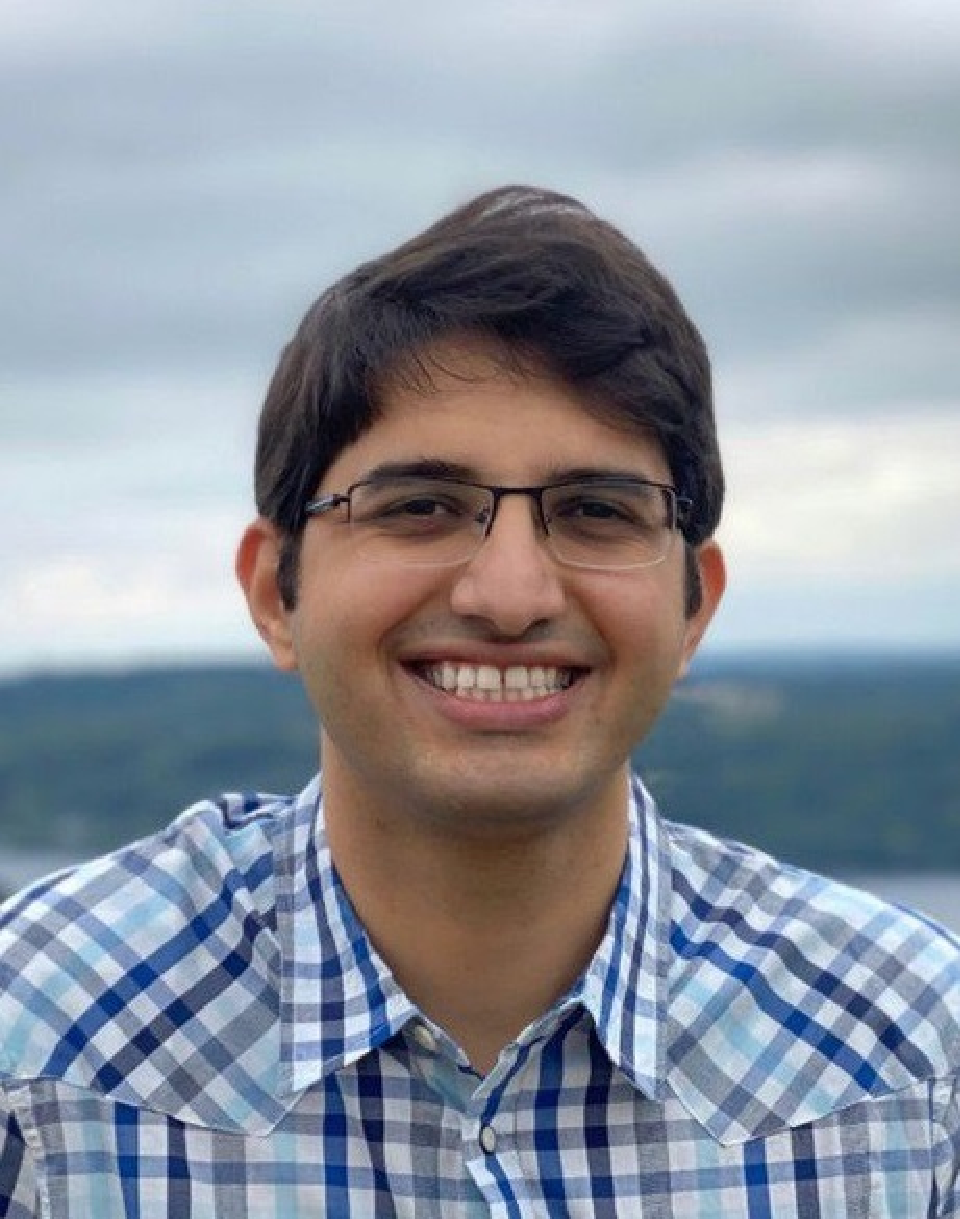}}]{SAJAD REZAIE} received his B.Sc. and M.Sc. degrees in Electrical Engineering from Amirkabir University of Technology, Iran, in 2013 and 2016, respectively. In 2023, he completed his Ph.D. in Wireless Communication at Aalborg University, Denmark. Since 2022, he has held the position of Senior Research Specialist at Nokia in Aalborg, Denmark, where he focuses on pioneering advancements in digital and wireless communications. His research interests span MIMO communications, mmWave communications, signal processing, and artificial intelligence applications in wireless systems. He has also contributed to several publications and patents, with samples of his work available at \url{https://github.com/SajadRezaie/}.
\end{IEEEbiography}

\begin{IEEEbiography}[{\includegraphics[width=1in,height=1.25in,clip,keepaspectratio]{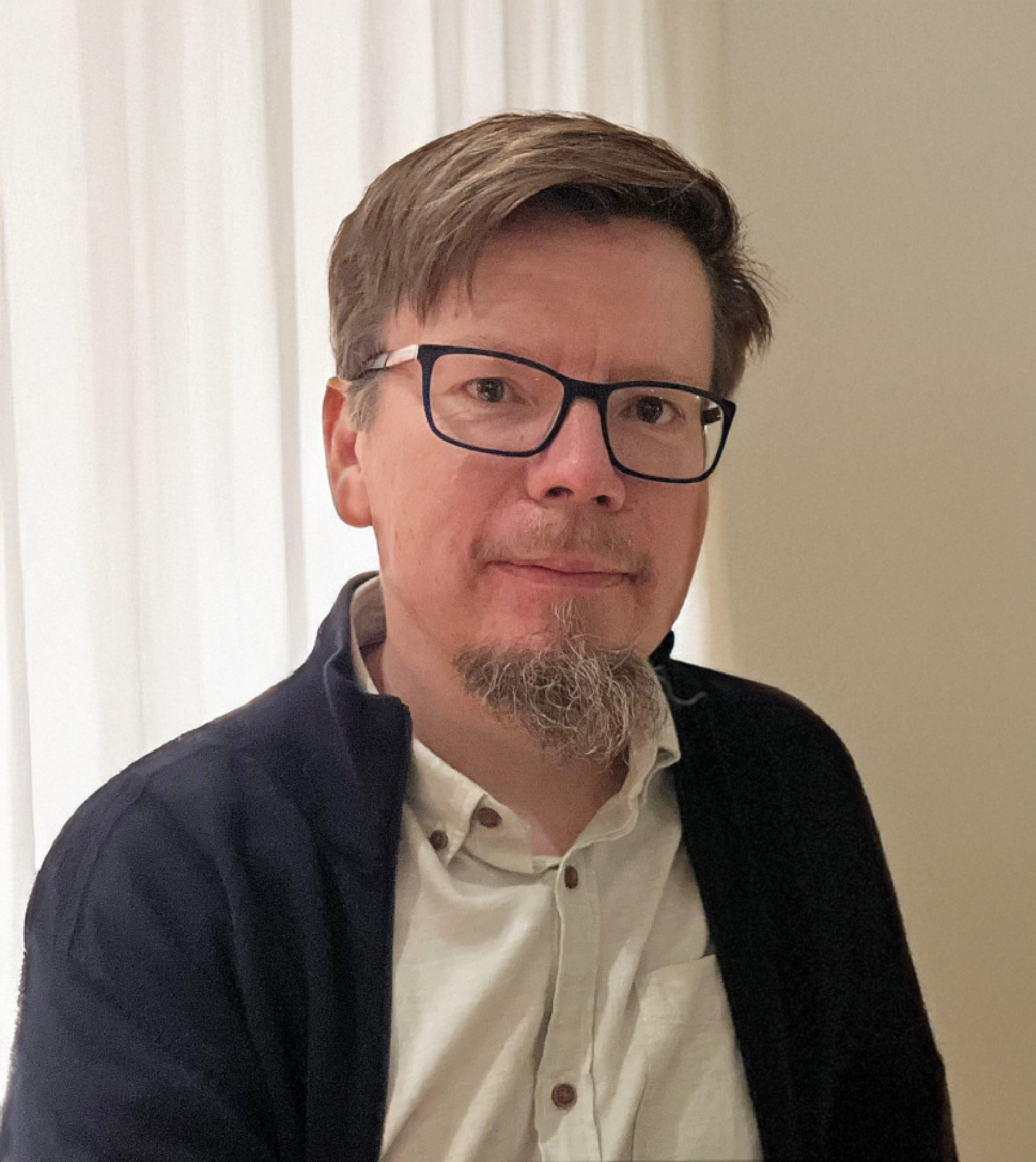}}]{MIKKO HONKALA} 
obtained his M.Sc. and Ph.D. degrees in Computer Science from Helsinki University of Technology (now part of Aalto University) in 2001 and 2007, respectively. He joined Nokia Research Center in 2008 and has since contributed to several advancements in the fields of machine learning and wireless communications, among others. He is currently leading a machine learning team at Nokia Bell Labs, Espoo, Finland. Some of his publications and patents are available at \url{https://scholar.google.fi/citations?user=Tw718yMAAAAJ}.
\end{IEEEbiography}

\begin{IEEEbiography}[{\includegraphics[width=1in,height=1.25in,clip,keepaspectratio]{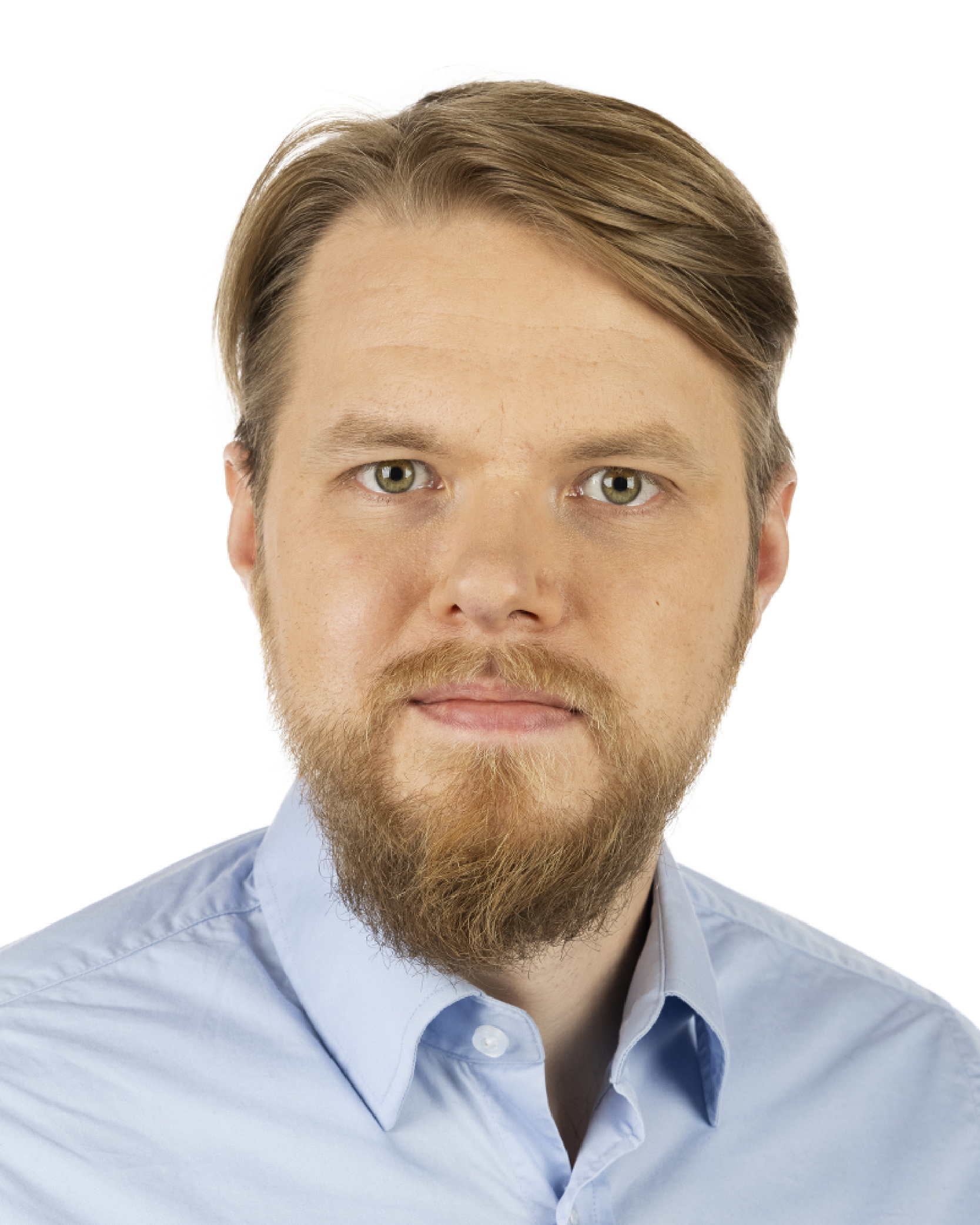}}]{DANI KORPI} 
received the D.Sc. degree (Hons.) in electrical engineering from Tampere University of Technology (TUT), Finland, in 2017. He is currently a Senior Specialist with Nokia Bell Labs, Espoo, Finland. His Ph.D. thesis received the Best Dissertation of the Year Award from TUT, as well as the Finnish Technical Sector’s Award for the best doctoral dissertation of 2017. His current research focuses on applying machine learning to wireless communications, especially on the physical layer. In particular, he is working on building a native foundation for machine learning in 6G radio networks.
\end{IEEEbiography}

\begin{IEEEbiography}[{\includegraphics[width=1in,height=1.25in,clip,keepaspectratio]{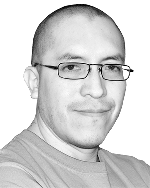}}]{DICK CARRILLO MELGAREJO} 
(M'06) received the B.Eng. degree (Hons.) in electronics and electrical engineering from San Marcos National University, Lima, Per\'u, and the M.Sc. degree in electrical engineering from Pontifical Catholic University of Rio de Janeiro, Rio de Janeiro, Brazil, in 2004 and 2008, respectively. 
Between 2008 and 2010, he contributed to WIMAX (IEEE 802.16m) standardization. 
From 2010 to 2018, he worked on the design and implementation of cognitive cellular radio networks based on 3GPP technologies. 
Since 2018 he is a researcher at Lappeenranta--Lahti University of Technology, where he received the D.Sc. degree in electrical engineering. 
Since 2022, he is a Senior Standardization Specialist at Nokia, where he is contributing on shaping the 3GPP standards toward 6G radio networks.
His research interests are mobile technologies 6G and ML applications in the radio interface.
\end{IEEEbiography}

\begin{IEEEbiography}[{\includegraphics[width=1in,height=1.25in,clip,keepaspectratio]{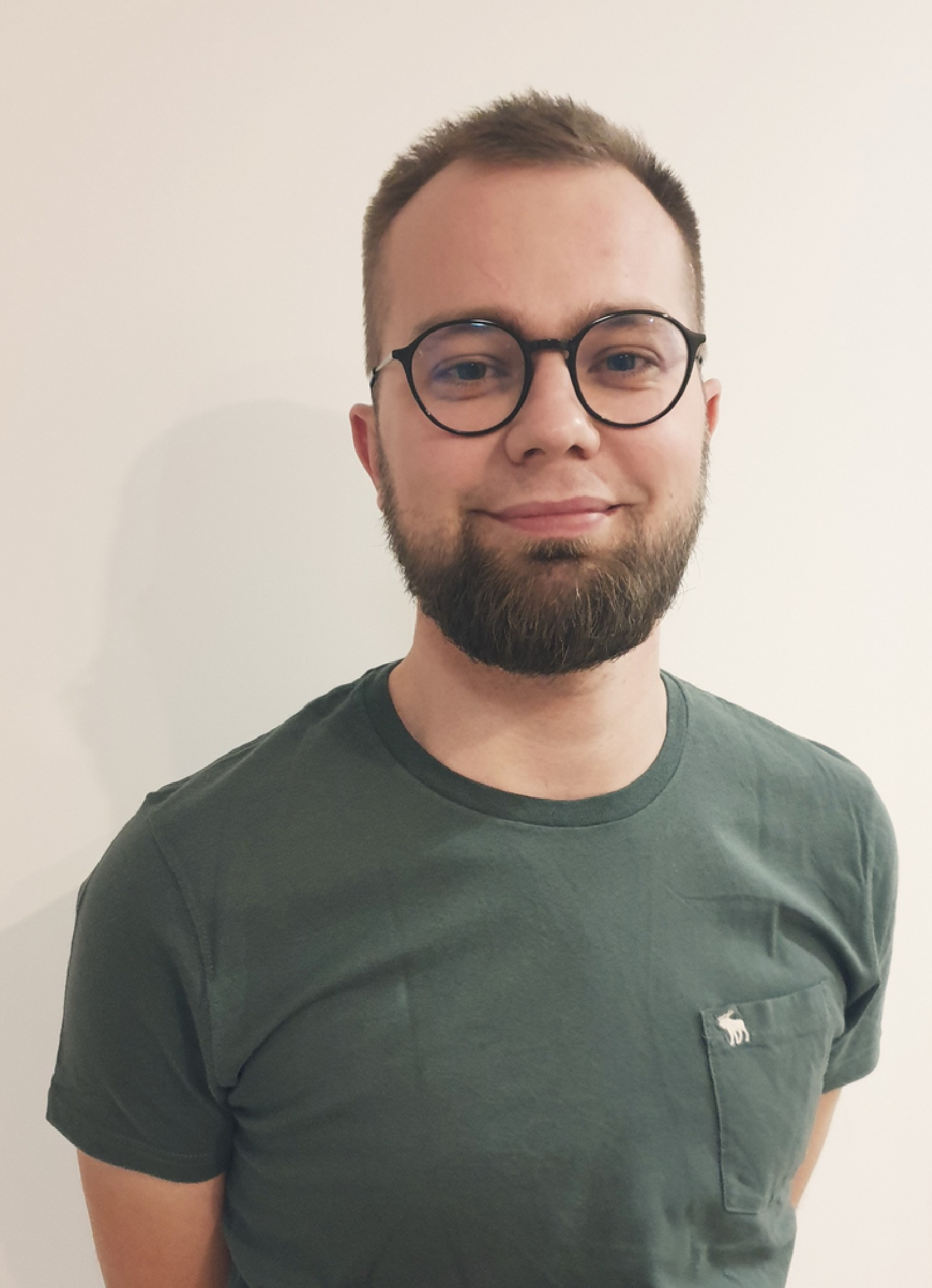}}]{TOMASZ IZYDORCZYK} 
received his M.Sc. in Wireless Communications Engineering from Eurecom, France. He completed his Ph.D. at Aalborg University, Denmark in 2020. Since 2019, he has held various research positions at Nokia in Wrocław, Poland, focusing on PHY and MAC design in 5G. In his current role as a RAN 2 delegate, his main research interests include generic 6G user plane design and artificial intelligence applications in wireless systems. He has also co-authored numerous publications and patent applications.
\end{IEEEbiography}

\begin{IEEEbiography}[{\includegraphics[width=1in,height=1.25in,clip,keepaspectratio]{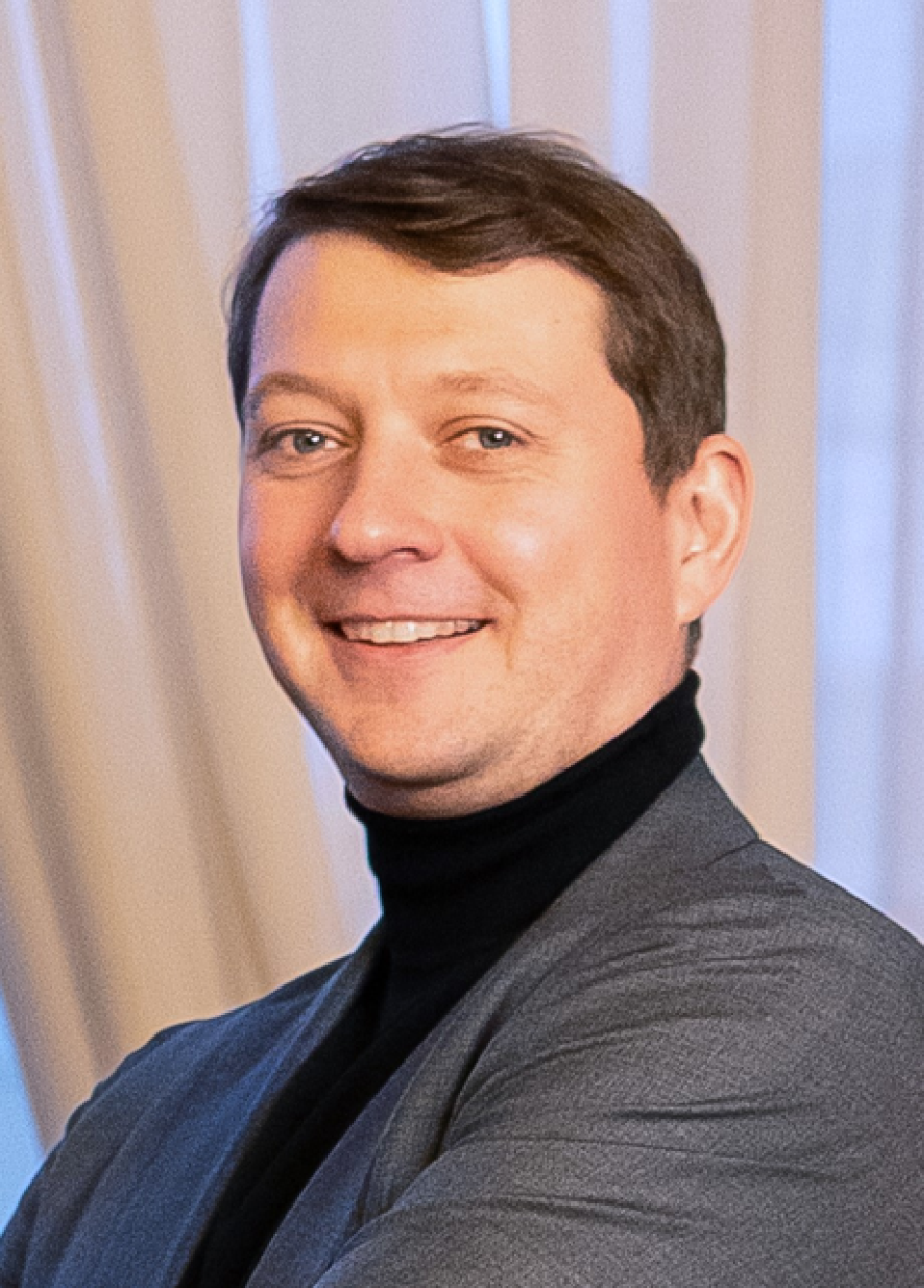}}]{DIMITRI GOLD} 
(Senior Member, IEEE) received the M.Sc. degree (2007, Hons.) in mathematical physics from M.V. Lomonosov Moscow State University, where he also pursued Ph.D. studies in mathematical modeling from 2007 to 2010. He received the Ph.D. degree (2012, excellent) and the title of Adjunct Professor (Docent, 2016) in mathematical information technology and telecommunications from the University of Jyväskylä, Finland. He has held senior research and leadership roles in both academia and industry, including at Nokia Bell Labs and Nokia Standards. Dr. Gold is an active contributor to 3GPP RAN1 and RAN4. His professional interests include future wireless access networks, machine learning in communications, and network modeling, bridging theoretical research with practical implementations.
\end{IEEEbiography}

\begin{IEEEbiography}[{\includegraphics[width=1in,height=1.25in,clip,keepaspectratio]{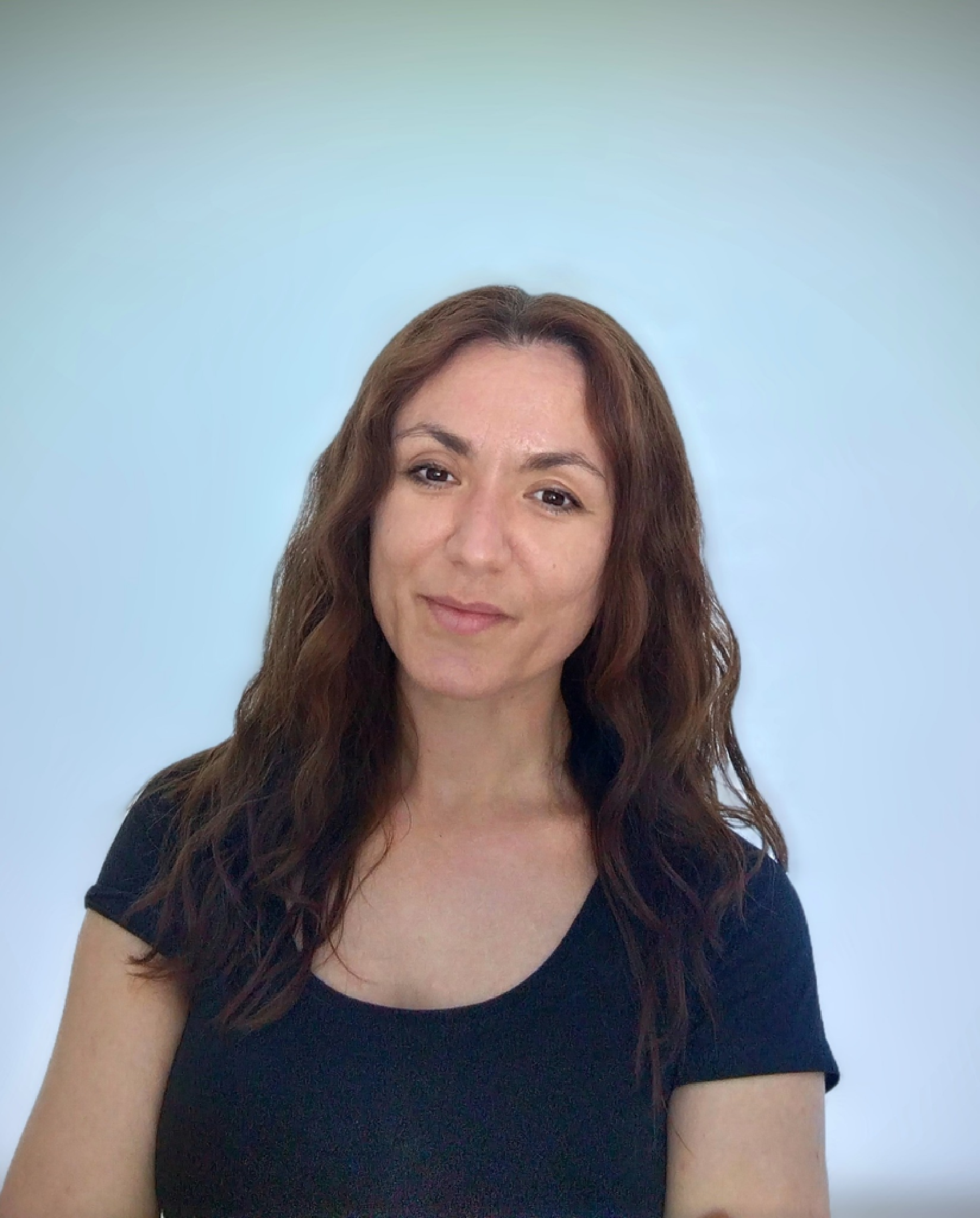}}]{OANA-ELENA BARBU} 
received the Ph.D. degree in wireless communications from Aalborg University, Denmark, in 2016. She is currently a Senior Staff Specialist with Nokia Denmark. Since 2016, she has held several research positions that focus on baseband receiver design using variational Bayesian inference and machine learning tools. Her current research focuses on machine learning solutions for advancing the physical layer design of 6G systems. Her co-authored publications and patents are available at \url{https://scholar.google.com/citations?user=TRSsDkkAAAAJ&hl=en&oi=ao}.
\end{IEEEbiography}

\end{document}